\documentclass[prd,superscriptaddress,showpacs,floatfix,%
preprintnumbers, nofootinbib,hyperref]{revtex4}              
\usepackage{graphicx}
\usepackage{rotating}
\usepackage{epsfig}
\usepackage[usenames,dvipsnames]{xcolor}

\newcommand{\mean}[1]{\langle #1 \rangle}
\usepackage{color}
\newcommand{\ovdmsolnu}{\overline{\Delta m^2_{{\odot},\nu}}}
\newcommand{\ovdmsolanu}{\overline{\Delta m^2_{{\odot},\bar\nu}}}
\begin{document}
\preprint{LAPTH-035/12, MPP-2012-117}

\title{
{\color{Orange} 
(Down-to-)Earth matter effect in supernova neutrinos 
}
}

\author{Enrico Borriello}

\author{Sovan Chakraborty} 

\author{Alessandro Mirizzi} 
\affiliation{II~Institut~f\"ur~Theoretische~Physik,~Universit\"at~Hamburg,~Luruper~Chaussee~149,~22761~Hamburg,~Germany}

\author{Pasquale Dario Serpico} 
\affiliation{LAPTh, Univ. de Savoie, CNRS, B.P.110, Annecy-le-Vieux F-74941, France}

\author{Irene Tamborra}
\affiliation{Max-Planck-Institut f\"ur Physik (Werner Heisenberg
Institut),
F\"ohringer Ring 6, 
80805 M\"unchen, Germany}


\begin{abstract}
Neutrino oscillations in the Earth matter may introduce peculiar modulations in the supernova (SN) neutrino spectra.
The detection of this effect has been proposed as diagnostic tool for the neutrino mass hierarchy at 
``large'' 1-3 leptonic mixing angle $\theta_{13}$. 
We perform an updated study on the observability of this effect at large next-generation underground detectors
(i.e., 0.4 Mton water Cherenkov, 50 kton scintillation and 100 kton liquid Argon detectors)  based on
neutrino fluxes from   state-of-the-art  SN simulations and accounting for statistical fluctuations via Montecarlo simulations. 
Since the  average energies predicted by recent simulations are  lower than 
previously expected and a tendency towards the  equalization of the neutrino fluxes appears 
during the SN cooling phase, the detection 
of the Earth matter effect  will be more challenging than expected from previous studies. We find that 
none of the proposed detectors shall be able to detect the Earth modulation for the neutrino signal of a typical galactic SN at
10~kpc. 
It  should be  observable  in a 100 kton  liquid Argon detector for a SN at few kpc and 
 all three detectors would clearly see the Earth signature for very close-by stars only ($d \sim 0.2$~kpc).
Finally,  we show that adopting IceCube as co-detector together with a Mton water 
Cherenkov detector is not a viable option either.

\end{abstract}


\maketitle

\section{Introduction}

Physical and astrophysical diagnostics via supernova (SN) neutrino detection in underground detectors
represent a subject of intense investigation in astroparticle physics. 
A lot of
attention has been devoted to possible signatures of the
Mikheyev-Smirnov-Wolfenstein (MSW) matter effect~\cite{Matt} 
on the neutrino flavor evolution in supernovae (SNe)~\cite{Dighe:1999bi}. 
Lately,  novel phenomena have
been found to be important in the region close to the
neutrinosphere where the neutrino density is  such that 
 the neutrino-neutrino interactions dominate
the flavor evolution. Neutrino self-interactions  are responsible for large coherent conversions
between different flavors (see, e.g.,~\cite{Duan:2010bg} for a recent review). 
 Neutrino oscillations in SNe  could imprint 
peculiar signatures on the observable neutrino signal, sensitive 
to  the neutrino
mass and mixing, and to the unknown  mass hierarchy~\cite{Raffelt:2012kt}. 
 
 Due to the uncertainties in the calculation of the primary SN neutrino spectra, 
it seems difficult to establish oscillation effects solely on the basis of
theoretical expectations. Therefore the importance of  relatively model-independent signatures has been 
emphasized in the recent literature,
e.g. in association with the prompt $\nu_e$ neutronization burst~\cite{Kachelriess:2004ds}, with the
rise time of the early neutrino signal~\cite{Serpico:2011ir}, or with matter effects associated to the shock-wave propagation
at late times~\cite{Schirato:2002tg,Fogli:2003dw,Fogli:2004ff,Fogli:2006xy,Tomas:2004gr}. One unequivocal signature would be the observation of the
Earth matter effects~\cite{Dighe:1999bi,Lunardini:2001pb,Dasgupta:2008my}. They induce a characteristic energy-dependent modulation
on the measured flux when SN neutrinos cross the Earth matter before being detected.
Earth matter effects could be measured in a single detector, if it has enough energy
resolution and statistics to track the wiggles in the observed energy spectrum. A Fourier analysis of the SN neutrino signal  has been
proposed as a powerful tool to diagnose this modulation~\cite{Dighe:2003jg,Dighe:2003vm}.
Moreover,  the comparison of the signals from 
shadowed and unshadowed detectors may allow one to diagnose the Earth effects even if 
a single detector could not resolve the modulations~\cite{Dighe:2003be}.

 Recent supernova simulations indicate lower average energies
than previously expected~\cite{Fischer:2009af,Huedepohl:2009wh,Serpico:2011ir} and a tendency towards the  equalization of the neutrino fluxes
during the cooling phase,  i.e. for post-bounce times $t \gtrsim 1$~s~\cite{Fischer:2011cy,Huedepohl:2009wh}. 
Remarkably,  this trend  eases the agreement of theoretical expectations with SN 1987A data~\cite{Jegerlehner:1996kx,Mirizzi:2005tg}.
Motivated by this new input and since the  observability of the Earth matter effect largely  depends on the neutrino  
average energies  and on the flavor-dependent differences among
the primary spectra, we find worthwhile to reevaluate the detectability of the Earth matter effects 
in large detectors. 
We refer to three classes of detectors proposed   for low-energy neutrino
physics and astrophysics, viz. water Cherenkov (WC) detectors
with fiducial masses of ${\cal O}$(400) kton~\cite{deBellefon:2006vq,Abe:2011ts}, liquid scintillation (SC) detectors with masses of
 ${\cal O}$(50) kton~\cite{Wurm:2011zn}, and Liquid Argon Time Projection Chambers
(LAr TPC) with fiducial masses of ${\cal O}$(100) kton~\cite{Badertscher:2010sy}. These three detection techniques are
the backbones of the European project LAGUNA (Large
Apparati for Grand Unification and Neutrino Astrophysics)~\cite{Autiero:2007zj}
 and the LBNE (Long Baseline Neutrino
Experiment) towards DUSEL (Deep Underground Science and
Technology Laboratory) in the US~\cite{Akiri:2011dv}.
Moreover, the project of  the Mton-class WC detector Hyper-Kamiokande
is currently discussed in Japan~\cite{Abe:2011ts}. 
In particular, WC and
SC neutrino experiments are mostly sensitive
to SN $\bar\nu_e$ through the inverse beta decay process
$\bar\nu_e+p\to n+e^+$. On the other hand, LAr TPC would have
a high sensitivity to SN $\nu_e$'s, through the charged current
interactions of $\nu_e$ with the Ar nuclei in the detector.
We also consider the  Icecube ice Cherenkov detector~\cite{Kopke:2011xb}  as ``co-detector''  to monitor the
Earth effect in comparison with the energy-integrated signal measured in a Mton WC 
detector~\cite{Dighe:2003be}. 

 For all these detectors, we  find that no signature of the Earth matter effect is observable 
for a typical galactic SN  at $d \simeq 10$~kpc. 
In the more  optimistic case of a  close-by SN at $d=1$~kpc, the chances
to detect the  Earth matter
signature appear statistically weak in the antineutrino signal. Conversely, a signal would show up
in the $\nu_e$ signal detectable at a LAr TPC. The Earth matter
signal would be detectable with high significance in both neutrino and antineutrino signals for relatively close by stars 
which might evolve into core-collapse SNe at unpredictable future times, like Betelgeuse, Mira Ceti and Antares (at $d\lesssim 0.2$~kpc) 
 --- provided that the electronics of the detector will be able to cope with such a huge rate of events.
Our results allow us to conclude that the previous paradigm on the observability of the 
SN neutrino Earth matter effects was based on an ``optimistic'' choice of the non-oscillated neutrino fluxes, not anymore confirmed by the 
most recent and less approximated SN simulations.
  
The plan of our work is as follow. In Sec.~II we present the neutrino signal   from  SN 
hydrodynamical simulations  we adopt
 as benchmark for our study. 
 In Sec.~III we characterize the SN neutrino flavor conversions and the Earth matter effect.
 In Sec.~IV  the features of our reference detectors (fiducial mass, cross sections and
 energy resolution) are described. In Sec.~V our results on  the detectability of the 
Earth effect in each of the reference  detectors are presented. Comments and conclusions are illustrated in
 Sec.~VI.

\section{Numerical models for supernova neutrino emission}

\begin{figure*}[!ht]
\begin{center}
\includegraphics[width=0.7\textwidth, angle=270]{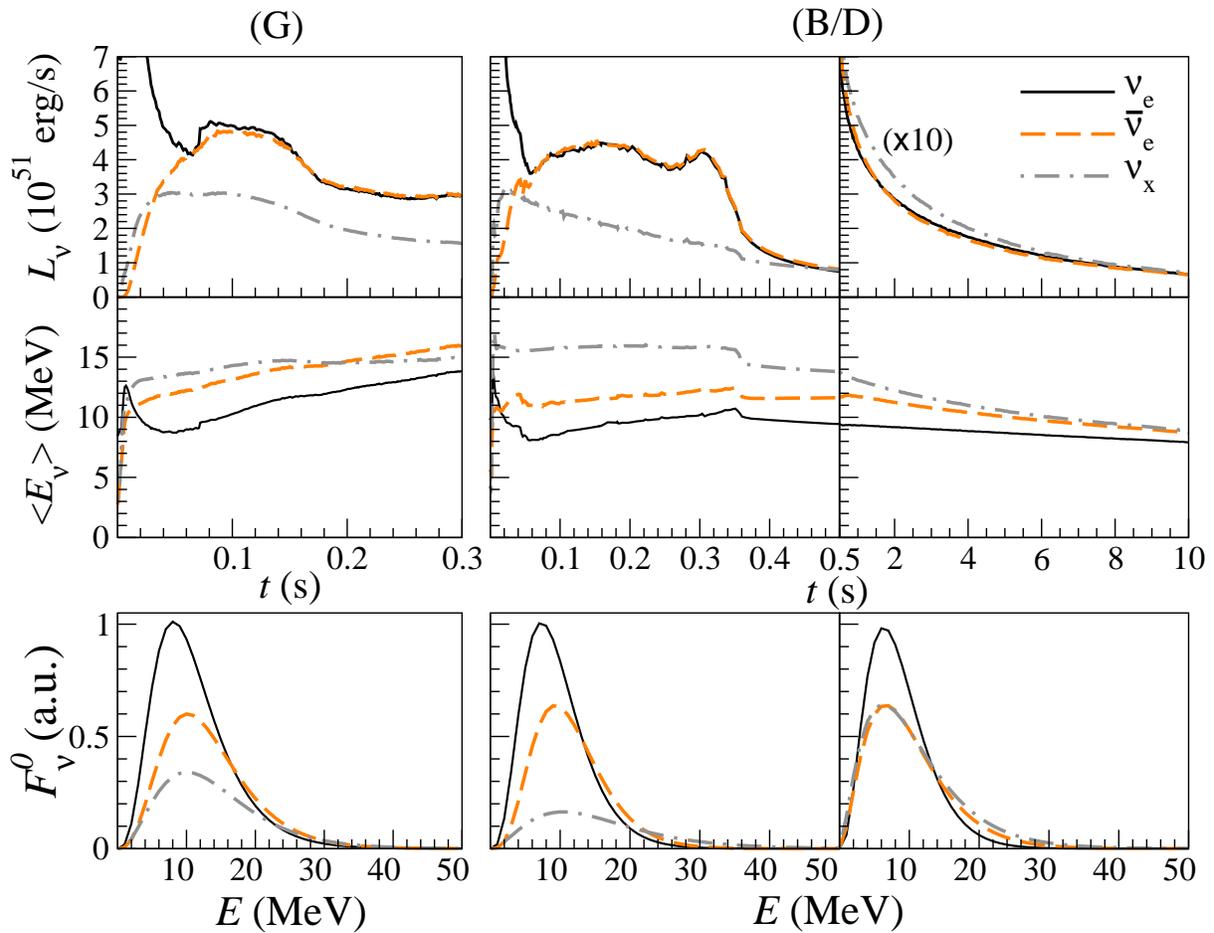}
\caption{Time evolution of neutrino luminosities $L_\nu$ (upper panels),  average energies
$\langle E_\nu \rangle$ (middle panels) and time-integrated energy spectra (lower panels)  for $\nu_e$ (continuous curve), $\bar\nu_e$ (dashed curve), and
$\nu_x$ (dash-dotted curve). The left panels refer to the accretion phase of a Garching simulation for a 15.0~$M_{\odot}$ progenitor, while
the other ones to a Basel/Darmstadt simulation for a  18.0~$M_{\odot}$ for the accretion (central panels) and
cooling (right panels) phase.}
\label{fig:lumen}
\end{center}
\end{figure*}

The un-oscillated $\nu$ distributions (with $\nu =\left\{ \nu_e, \overline\nu_e, \nu_x\right(=\nu_\mu , \nu_\tau) \}$) 
are parametrized in energy and time as follows
\begin{equation}
F^0_\nu (E,t) =\phi_\nu(t)\, f_\nu(E,t) = \frac{L_\nu(t)}{\langle E_\nu(t)\rangle} f_\nu(E,t) \,\ ,
\end{equation}
where $\phi_{\nu}(t)$ is the  \emph{energy-integrated} neutrino number flux for each  post-bounce time $t$, $L_\nu (t)$ the luminosity and 
 $\langle E_\nu(t)\rangle$ the mean neutrino energy. The function $f_\nu(E,t)$ is the energy spectrum, normalized such that $\int dE\ f_\nu(E,t) = 1$. 
 It is parametrized as in~\cite{Keil:2002in}
\begin{equation}
f_\nu(E,t)= \frac{1}{\langle E_\nu(t)\rangle}\frac{(1+\alpha_\nu(t))^{1+\alpha_\nu(t)}}{\Gamma(1+\alpha_\nu(t))}\left(\frac{E}{\langle E_\nu(t)\rangle}\right)^{\alpha_\nu(t)}\exp\left[-(1+\alpha_\nu(t))\frac{E}{\langle
E_\nu(t)\rangle}\right]\, ,
 \label{eq:varphi}
\end{equation}
where the energy-shape parameter $\alpha_\nu(t)$ is~\cite{Keil:2002in,Raffelt:2001}
\begin{equation}
\alpha_\nu(t)=\frac{2\langle E_\nu(t) \rangle^2-\langle E_\nu(t)^2\rangle}{\langle E_\nu(t)^2\rangle-
\langle E_\nu(t)\rangle^2} \, . \label{alphadef}
\end{equation} 
The variables  $L_\nu(t)$, $\langle E_\nu(t)\rangle$ and $\alpha_\nu(t)$ are  extracted from SN simulations.

 In the following we adopt as input the  recent SN  simulations performed
by the Garching group (G) ---see~\cite{Serpico:2011ir} for references--- and by the Basel/Darmstadt group (B/D)~\cite{Fischer:2009af}.  
In Fig.~\ref{fig:lumen} we show the initial luminosities $L_{\nu}$ (upper panels), average energies $\langle E_\nu \rangle$ (middle panels) and time-integrated energy-spectra in different time windows (lower panels) for the different flavors $\nu_e$, $\bar\nu_e$ and $\nu_x$ and for the simulations from the two groups:
i) accretion phase of a 15 $M_{\odot}$ progenitor from the Garching group  (left panel); ii) accretion and cooling phases of a 18 $M_{\odot}$ progenitor  from the Basel/Darmstadt group (middle panels for the accretion and right panels for the cooling).

In the former case, the accretion phase is clearly visible in the neutrino light-curve. 
It lasts till $t \simeq 0.2$~s and appears as a pronounced hump in the electron
(anti)neutrino luminosities.
 The relative time-integrated spectra for the accretion phase (with $t \in 
[0,0.25]$~s) show that flavor-dependent flux differences are large, with a robust
hierarchy for the time-integrated neutrino number fluxes
($\Phi_{\nu}^0=\int dt \,\ \phi_{\nu}(t)$), namely $\Phi_{\nu_e}^0 > \Phi_{{\overline\nu}_e}^0 \gg \Phi_{\nu_x}^0$. 
Therefore, the accretion phase would represent the best time-window to detect effects of flavor conversions.

The Basel/Darmstadt simulations have an accretion phase lasting till $t \simeq 0.4$~s (middle panel).  
 The main features of the neutrino signal are similar to the ones of the Garching case.
These simulations provide  the late-time SN neutrino signal from the cooling phase too,  
shown in the right panels of Fig.~\ref{fig:lumen}. 
During the cooling phase all neutrino flavors originate
close to the neutron star surface, where the material is  neutron rich, suppressing charged-current reactions for ${\overline\nu}_e$. Therefore, 
 the luminosities and the average energies of ${\overline\nu}_e$ and $\overline{{\nu}}_x$ become quite similar. As a result,  the time-integrated  neutrino fluxes in 
$t \in [1.0,10.0]$~s are so close that one expects to be difficult to measure
oscillation effects in this case. 

In Table~\ref{tab:fluxes} we report the parameters of the  time-integrated SN neutrino spectra
 for the Garching  and the Basel/Darmstadt simulations distinguishing between the accretion ($t \le 0.25$~s for G and $t \le 0.4$~s for B/D) 
 and the cooling phase. These parameters will be taken as 
benchmark for the evaluation of the Earth matter effect.  
 From this Table, it is evident that the neutrino average energies provided by the recent SN simulations
are significantly lower than what assumed in the previous studies (see, e.g. Table 3 of~\cite{Dighe:2003jg}).

\begin{table}[tdp]
\caption{Spectral-fit parameters for the neutrino and antineutrino
fluxes  integrated over  the accretion phase of the 15 $M_{\odot}$ Garching
progenitor (G) and both the accretion and cooling phases of the 18
$M_{\odot}$   Basel/Darmstadt (B/D) progenitor.}
\begin{center}

\begin{tabular}{ccccccc}
\hline
 Model &  $\mean{E_{\nu_e}}$ (MeV)  & $\mean{E_{\nu_x}}$
(MeV) & $\Phi_{\nu_e}^0 (\times 10^{56})$ & $\Phi_{\nu_x}^0 (\times 10^{56})$ & ${\alpha}_{\nu_e}$ &  ${\alpha}_{\nu_x}$ \\
\hline \hline
  G (accretion, $t \le 0.25$~s)   &  10.9 & 14.0 & 5.68 & 2.67 & 3.1 & 2.5 \\
  B/D (accretion, $t \le 0.4$~s)  &  9.5 & 15.6 & 8.53 & 3.13 & 3.4 & 2.0 \\
  B/D (cooling, $t > 1.0$~s)  &  8.6 & 10.5 & 11.80 & 10.75 & 2.8 & 1.5 \\
\hline\\
\end{tabular}

\begin{tabular}{ccccccc}
\hline
 Model &  $\mean{E_{\bar\nu_e}}$ (MeV)  &
$\mean{E_{\bar\nu_x}}$ (MeV)  & $\Phi_{\bar\nu_e}^0$ $(\times10^{56}$) &
$\Phi_{\bar\nu_x}^0 $ $(\times10^{56}$) & ${\alpha}_{\bar\nu_e}$ &
${\alpha}_{\bar\nu_x}$   \\ \hline \hline
  G (accretion, $t \le 0.25$~s)   &  13.2 & 14.0 & 4.11 & 2.67 & 3.3 & 2.5 \\
  B/D (accretion, $t \le 0.4$~s)  &  11.6 & 15.6 & 7.51 & 3.13 & 4.0 & 2.0 \\
  B/D (cooling, $t > 1.0$~s)  &  10.0 & 10.5 & 9.74 & 10.75 & 1.9 & 1.5 \\
\hline
\end{tabular}
\end{center}
\label{tab:fluxes}
\end{table}
%

\section{Flavor conversions and Earth matter effect}

\subsection{Neutrino mixing parameters}

We assume the three neutrino mass eigenstates separated by the following neutrino mass squared differences as from the global $3\ \nu$ oscillation analysis~\cite{Fogli:2012ua}  
\begin{eqnarray}
\label{masses}
\Delta m^2_{\rm atm} &=& m_3^2 - m_{1,2}^2 = 2.35\times 10^{-3}\mathrm{\ eV}^2\ ,\\
\Delta m^2_{\odot} &=& m_{2}^2-m_{1}^2 = 7.54\times 10^{-5}\mathrm{\ eV}^2\ .
\end{eqnarray}
Since the sign of $\Delta m_{\rm atm}^2$ is not determined yet,  we will consider both normal (NH, $\Delta m_{\rm atm}^2>0$) and inverted hierarchy (IH, $\Delta m_{\rm atm}^2<0$) scenarios. 
The mass eigenstates are related to the flavor eigenstates $(\nu_e,\nu_\mu,\nu_\tau)$ by means of three mixing angles. 
Their best-fit values, as in the global $3\ \nu$ oscillation analysis in~\cite{Fogli:2012ua}, are
\begin{eqnarray}
\label{theta1312}
\sin^2\theta_{13}=0.02\quad \mathrm{and}\quad \sin^2\theta_{12}=0.31\ .
\end{eqnarray}
The mixing angle $\theta_{23}$ is not relevant for our purposes since we are assuming
equal $\nu_\mu$ and $\nu_\tau$ fluxes.

\subsection{No Earth crossing}

The emitted SN neutrino flux is processed by self-induced and
MSW oscillation effects during its propagation. 
The self-induced effects would take place within $r \sim \mathcal{O}(10^{3})$~km from the neutrinosphere
whereas the MSW transitions take place at larger radii, in the region $r \sim 10^{4}$--$10^{5}$~km. 
As the self-induced and MSW effects are widely separated in space, they can be considered
independently of each other.

We start considering the accretion phase.
No self-induced flavor conversion   occurs in NH and for the spectral ordering of the accretion phase~\cite{Fogli:2007bk,Fogli:2008pt,Mirizzi:2010uz}.
Instead, potentially large self-induced effects could be expected for neutrinos and antineutrinos in IH~\cite{Fogli:2007bk,Fogli:2008pt,Mirizzi:2010uz}. 
However, it has been shown using results both  from Basel/Darmstadt group simulations and Garching group ones~\cite{Chakraborty:2011nf,Chakraborty:2011gd,Saviano:2012yh,Sarikas:2011am,Sarikas:2012vb} that the multi-angle effects associated with the dense ordinary matter suppress collective oscillations in iron-core SNe~\cite{EstebanPretel:2008ni}.

The neutrino fluxes can only undergo the  traditional MSW conversions
in SNe while passing through the outer layers of the star.
Therefore,  it is straightforward  to compute the
$\overline\nu_e$ flux at  the Earth in the different cases~\cite{Dighe:1999bi}. 
Recently reactor experiments measured  ``large''   $\theta_{13}$~\cite{An:2012eh,Ahn:2012nd},
for such  value of the mixing angle and in NH, one finds
\begin{eqnarray}
F_{\bar\nu_e} = \cos^2\theta_{12} F^0_{\bar\nu_e} + \sin^2\theta_{12} 
 F^0_{\bar\nu_x}\quad \ \  {\mathrm{and}}\quad \ \ F_{\nu_e} = F^0_{\nu_x}\ .
\label{eq:nh} 
\end{eqnarray}
Instead, in  IH   one gets
\begin{eqnarray}
F_{\bar\nu_e} = F^0_{\bar\nu_x} \quad \ \  {\mathrm{and}}\quad \ \ 
F_{\nu_e} = \sin^2\theta_{12}  F^0_{\nu_e} +  
\cos^2\theta_{12} F^0_{\nu_x} \,\ .
\label{eq:ih}
\end{eqnarray}

Concerning the cooling phase, the characterization of the flavor conversions is less straightforward. Indeed, it is expected
that self-induced flavor oscillations would be  no longer inhibited by the   ordinary matter density  and multiple spectral splits should occur~\cite{Dasgupta:2009mg,Fogli:2009rd}. However,
  since the spectral differences among different flavors are not large 
(see Fig.~\ref{fig:lumen}), we numerically checked that the effect of the self-induced oscillations would produce a flavor equilibration among the different neutrino species (see also~\cite{Mirizzi:2010uz,Lunardini:2012ne}). This would reduce the possibility to observe
any signature of Earth matter effects. Moreover, it is expected that at $t \gtrsim 2$~s the non-adiabatic
effects associated with  the   matter turbulences in the supernova envelope would produce a smearing of the MSW
flavor conversions~\cite{Fogli:2006xy,Friedland:2006ta,Kneller:2010sc} further  reducing the spectral differences. 
In the following, we will neglect all these complicated effects and we will assume that 
the oscillated fluxes are described by
 Eqs.~(\ref{eq:nh})--(\ref{eq:ih})  during the cooling phase too, since we are interested in the time-integrated 
signal. 
Our approach is conservative, since the experimental detectability of the Earth matter signature
would be even more challenging than in our simplified scenario.

\subsection{Earth crossing}

If the supernova is shadowed by the Earth for a detector~\cite{Mirizzi:2006xx}, 
 neutrinos will travel a certain distance through the Earth 
and therefore will undergo Earth matter oscillations during their propagation.
Since neutrinos arrive at the Earth as mass eigenstates,
the net effect of oscillations can be written 
in terms of the conversion probabilities $P_{ie}=P(\nu_i \to \nu_e)$. 
 For large $\theta_{13}$,  the neutrino fluxes at the Earth are~\cite{Dighe:1999bi}
\begin{equation} 
F_{\nu_e}^{\oplus}  = (1-P_{2e}) F^0_{\nu_e}  +  P_{2e} F^0_{\nu_x} \,\ 
\label{eme-nu}  
\end{equation}  
and for antineutrinos
\begin{equation} 
F_{\bar\nu_e}^{\oplus} =(1 - \bar{P}_{2e}) F^0_{\bar\nu_e}+
\bar{P}_{2e} F^0_{\bar\nu_x} \,\,
\label{eme-nubar}  
\end{equation}  
 Here $P_{2e} \equiv P(\nu_2 \to \nu_e)$  and $\bar{P}_{2e} \equiv P(\bar\nu_2 \to \bar\nu_e)$ 
while propagating through the Earth.
The analytical expressions for $P_{2e}$ and $\bar{P}_{2e}$ can be calculated
for the approximate two-density model of the Earth~\cite{Dighe:2003vm}.
When neutrinos traverse a distance $L$ through the mantle of the Earth,
these quantities  assume a very simple form~\cite{Dighe:1999bi,Lunardini:2001pb}:
\begin{eqnarray}  
P_{2e} & = & \sin^2\theta_{12} + \sin2\theta^m_{12} \, \label{P2e}
 \sin(2\theta^m_{12}-2\theta_{12})  
\sin^2\left(  
\frac{\Delta m^2_{\odot} \sin2\theta_{12}}{4 E \,\sin2\theta^m_{12}}\,L  
\right)\,, 
\\
\bar{P}_{2e} & = & \sin^2\theta_{12} + \sin2\bar\theta^m_{12} \, \label{Pbar2e}  
 \sin(2\bar\theta^m_{12}-2\theta_{12})  
\sin^2\left(  
\frac{\Delta m^2_{\odot}\,\sin2\theta_{12}}{4 E \,\sin2\bar\theta^m_{12}}\,L  
\right)\,, 
\end{eqnarray}  
where $\theta^m_{12}$ and $\bar\theta^m_{12}$ are the effective
values of $\theta_{12}$ in the Earth matter for neutrinos and antineutrinos, 
respectively~\cite{Fogli:2001pm}. 
The Earth crossing  induces a peculiar oscillatory signature in   
the energy spectra and from Eqs.~(\ref{eq:nh})--(\ref{eme-nubar}),  one would expect the Earth matter effect 
for antineutrinos in NH and for neutrinos in IH.

\subsection{Power spectrum of the Earth matter signal}

 The typical event rate associated with an inverse beta decay process  is $ \propto E^2 F_{\bar\nu_e}^{\oplus} (E)$,
 the cross section being $\sigma \propto E^2$.
 Whereas the distance between the energy-spectrum peaks  due to the Earth modulation increases with energy, 
the peaks  are nearly equally spaced in the inverse energy spectrum. 
In order to show this behavior,  it is useful to recast 
the $\nu_e$ and $\bar\nu_e$   fluxes 
as~\cite{Dighe:2003jg}
\begin{eqnarray}
F_{\nu_e}^{\oplus} &=& \sin^2\theta_{12}  F^0_{\nu_e} +  
\cos^2\theta_{12} F^0_{\nu_x} + \Delta {F}^0 {A}_{\oplus} \sin^2 \left (\frac{\ovdmsolnu}{10^{-5}\,{\rm eV}^2} \frac{L}{10^3\,{\rm km}} y\right) \,\ ,
\label{nueearth}\\
F_{\bar\nu_e}^{\oplus} &=& \cos^2\theta_{12} F^0_{\bar\nu_e} + \sin^2\theta_{12}
 F^0_{\bar\nu_x} - \Delta {\bar F}^0 \bar{A}_{\oplus} \sin^2 \left (\frac{\ovdmsolanu}{10^{-5}\,{\rm eV}^2} \frac{L}{10^3\,{\rm km}} y\right) \,\ ,
 \label{anueearth}
\end{eqnarray}
where $\Delta F^0 = F^0_{\nu_e} - F^0_{\nu_x}$ is the energy-dependent difference of the primary neutrino spectra and  ${A}_{\oplus} =  \sin(2\theta^m_{12}-2\theta_{12})$;
we also defined the mass squared difference in the Earth  $\ovdmsolnu =
\Delta m^2_{\odot} \sin2\theta_{12}/{\sin2 \theta^m_{12}}$ and the ``inverse energy variable''  
$y = 12.5\,{\rm MeV}/E$. 
Analogous definitions hold in the antineutrino sector in terms of the difference of the antineutrino spectra and of $\bar\theta^m_{12}$. 
The Earth matter effects on SN neutrino signal can thus be identified at a single detector through 
peaks in the Fourier transform of the inverse energy spectrum~\cite{Dighe:2003jg}. 
The neutrino signal is observed as a discrete set of events. 
Following~\cite{Dighe:2003jg}, we define the power spectrum of the $N$ detected events as
\begin{equation}
G_N (k) \equiv \frac{1}{N} \left| \sum_{i=1}^{N}e^{i k y_i} \right|^2 \,\ .
\label{power}
\end{equation}
This function is related to the continuous power spectrum, $G(k)$, by means of the function $q(y) \propto E^2 F_{\bar\nu_e}^{\oplus} (E)$:
\begin{equation}
G_N(k) = N \left|\frac{1}{N}\sum_{i=1}^{N}e^{i k y_i}\right|^2
= N  \left|\langle e^{iky} \rangle\right|^2 \approx
N \left|\int dy\,\ q(y) e^{iky} \right|^2 \equiv N G(k) \,\ .
\label{eq:contpower}
\end{equation}
 
%
\begin{figure*}[!t]
\begin{center}
\includegraphics[width=0.48\textwidth]{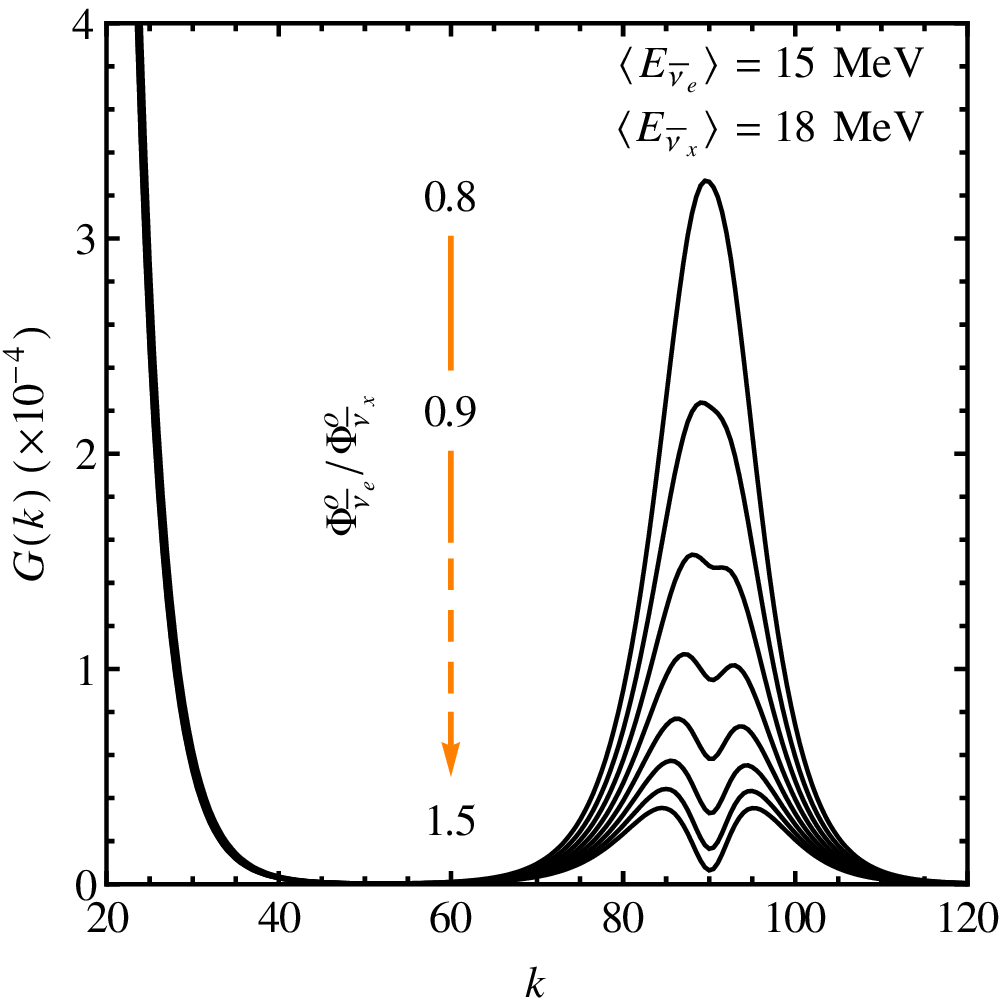}
\includegraphics[width=0.48\textwidth]{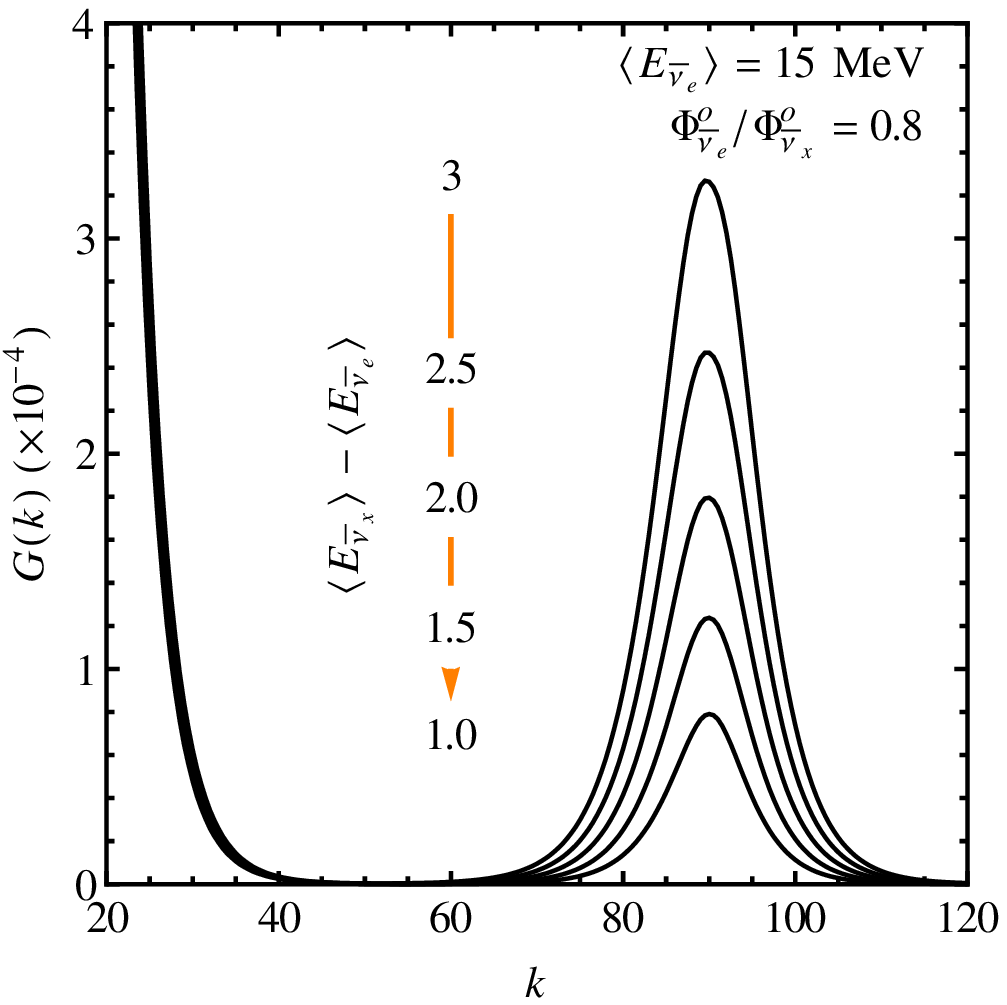}
\caption{Power spectrum $G(k)$ of the Earth matter signal. The left panel shows $G(k)$
for a flux ratio varying  in the range $0.8 < \Phi_{\bar\nu_e}^0/\Phi_{\nu_x}^0<1.5$,
while the right panel shows the impact of the change of the difference of the average energies 
in the range $1 \,\ \textrm{MeV}< \langle E_{\bar\nu_e} \rangle - \langle E_{\bar\nu_x} \rangle<3$~MeV.}
\label{fig:parametric}
\end{center}
\end{figure*}

Figure~\ref{fig:parametric} shows the power spectrum  $G(k)$ of the function 
$q(y)$ as a function of $k$,  for an  illustrative purpose. Note that the Earth effects introduce peaks in the power  spectrum at specific
frequencies. In particular, we aim at discussing the dependence of  $G(k)$  on the ratio of the 
initial fluxes $\Phi_{\bar\nu_e}^0/\Phi_{\bar\nu_x}^0$  (left panels), and on the difference among the  neutrino average energies 
$\langle E_{\bar\nu_x} \rangle -\langle E_{\bar\nu_e} \rangle$ (right panels). 
We take as  benchmark values the ones used in~\cite{Dighe:2003jg}, that refer to previous
Garching simulations~\cite{Raffelt:2003en} and
we fix 
$\langle E_{\bar\nu_e} \rangle = 15$~MeV and  $\alpha_{\bar\nu_e}=\alpha_{\bar\nu_x}=3.0$.  In the left panel, we assume
 $\langle E_{\nu_x} \rangle = 18$~MeV and 
vary the ratio of the total fluxes $\Phi_{\bar\nu_e}^0/\Phi_{\bar\nu_x}^0$ between 0.8 and 1.5. 
The two extreme cases are representative of the  flux ordering during the cooling and  the accretion
phase, respectively.
  The large peak in $G(k)$  at low values of $k$ is the dominant contribution due to the first two terms in 
Eq.~(\ref{anueearth}).   The peak at $k \simeq 90$ corresponds to the oscillations in the Earth
matter.  Since $\bar\nu_e$'s have lower average energy than $\bar\nu_x$'s, and due to the suppression at low-energy
associated with the $E^2$ dependence in the cross section, 
 the peak decreases significantly increasing the ratio $\Phi_{\bar\nu_e}^0/\Phi_{\bar\nu_x}^0$. 
 In the right panel, instead, we fix $\Phi_{\bar\nu_e}^0/\Phi_{\bar\nu_x}^0 =0.8$ and vary
  $\langle E_{\bar\nu_x} \rangle$ between 15 and 18~MeV. 
  As expected, since reducing the difference among the average energies,
$E^2\Delta {\bar F}^0 $ becomes smaller, the peak in the power spectrum is strongly suppressed.

From this parametric study, we expect that  
the detection of the Earth peak would be more challenging than what reported in the previous literature
for the SN models described in Sec.~II. 
 In fact the trend shown in Fig.~\ref{fig:parametric} is also illustrative of the expected differences in the power 
spectrum adopting old and more recent supernova simulation inputs. The average energies that we adopt for different $\nu$  species 
are lower and closer among themselves   than previously assumed. Therefore, we expect that the recent data, both for the accretion 
and the cooling phase, would produce a power spectrum with a significant suppression of the expected peak,
similar to the one with the smallest difference $\langle E_{\bar\nu_x} \rangle-\langle E_{\bar\nu_e} \rangle$ (see right panel of Fig.~\ref{fig:parametric}).
 We will present a quantitative estimation of this effect in the following Sections.

\begin{figure}[!t]
\includegraphics[width=.4\textwidth]{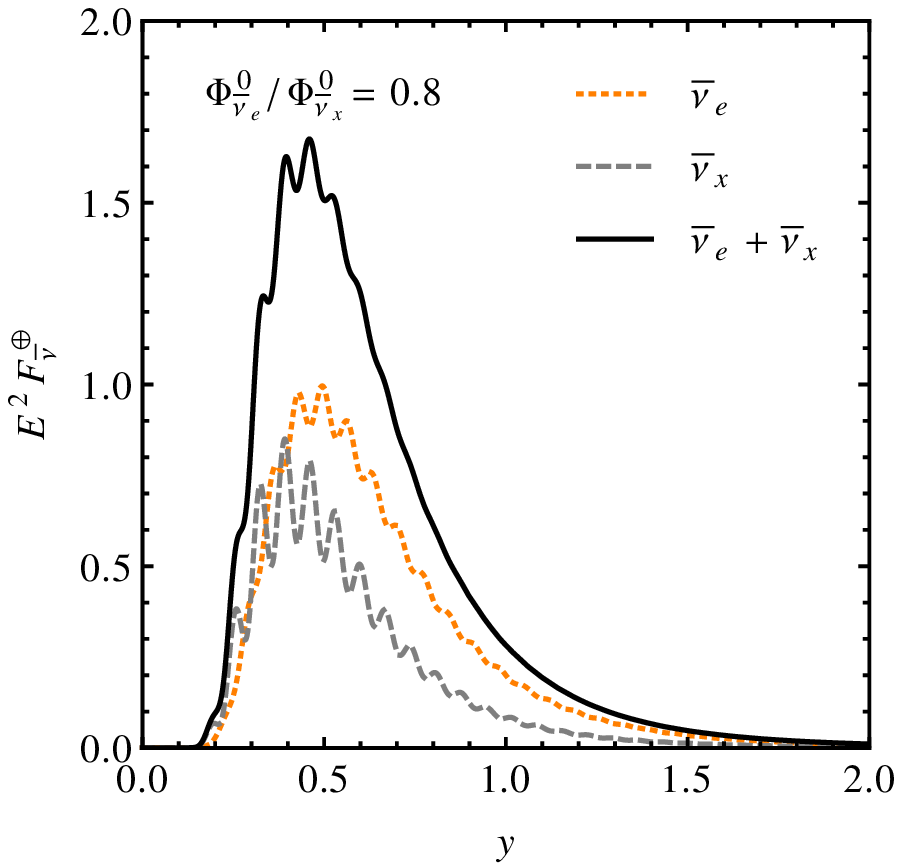}
\includegraphics[width=.414\textwidth]{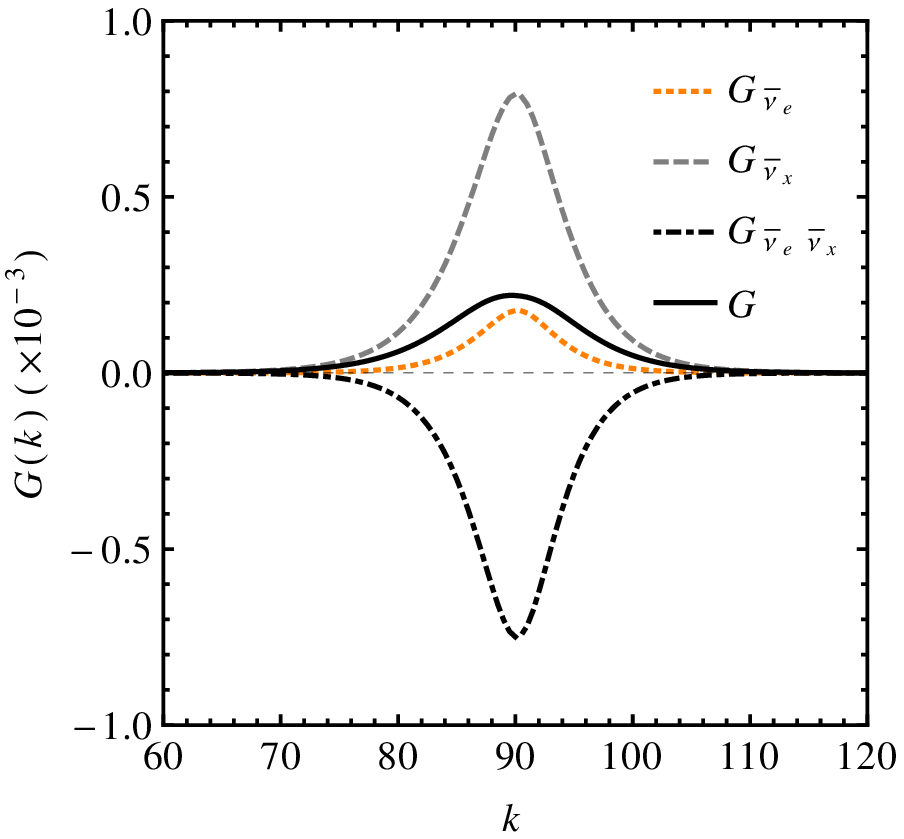}
\includegraphics[width=.4\textwidth]{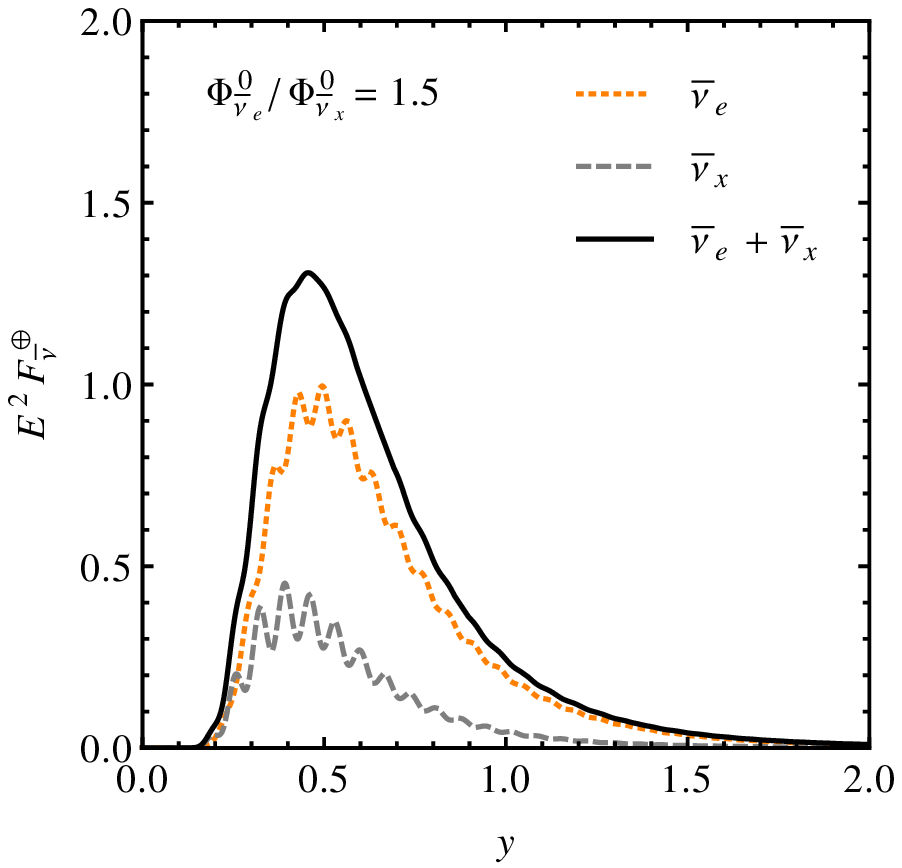}
\includegraphics[width=.414\textwidth]{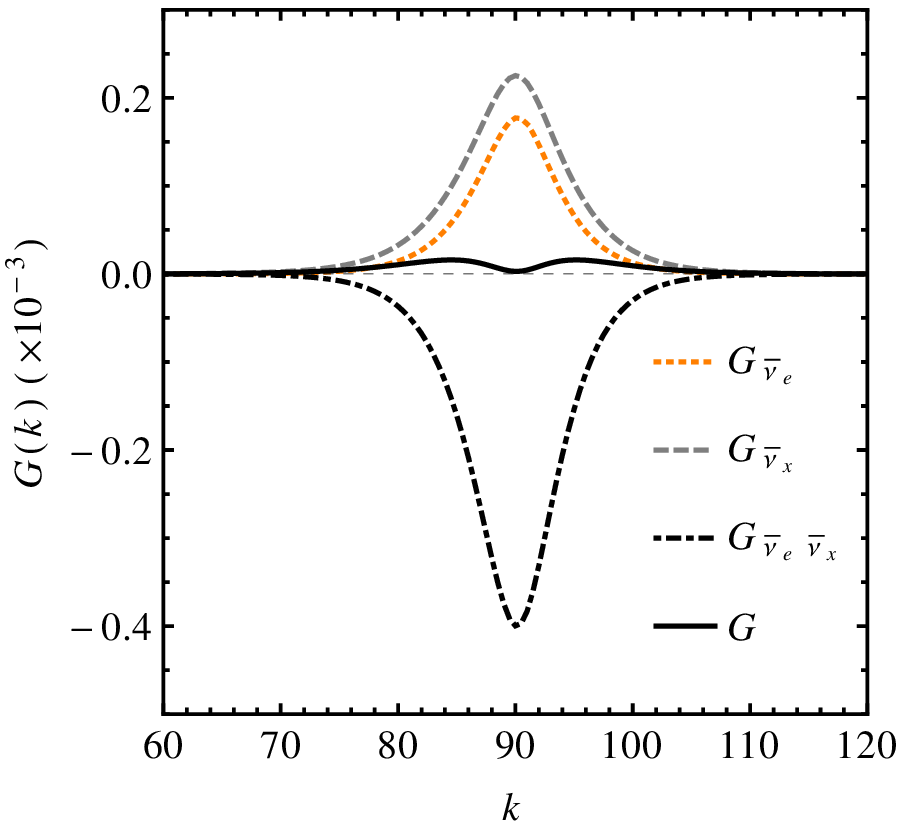}
\caption{Left panels: Contributions of the different flavors to the observable Earth-modulated signal  $E^2 F_{\bar\nu_e}^{\oplus}$, 
corresponding to the uppermost and lowermost cases of left panel in Fig.~\ref{fig:parametric}. 
 The continuous curve represents  the total   $E^2 F^{\oplus}_{\bar\nu_e}$ flux, the short-dashed
one  corresponds to $E^2 (1 - \bar{P}_{2e}) F^0_{\bar\nu_e}$, while the long-dashed one to
$E^2 \bar{P}_{2e} F^0_{\bar\nu_x}$ (see Eq.~\ref{anueearth}).
Right panels: Contributions to the power-spectrum $G(k)$ as from Eq.~\ref{eq:powerthree}.
 The continuous curve represents  the total  power-spectrum $G(k)$, 
 the short-dashed one corresponds to $G_{\bar\nu_e}(k)$, 
 the long-dashed one corresponds to $G_{\bar\nu_x}(k)$, while the dashed-dotted curve is for
 $G_{\bar\nu_e \bar\nu_x}(k)$
(see text for details).}
\label{fig:double-peak}
\end{figure}

We now comment on the appearance of the double-peak feature as the ratio $\Phi_{\bar\nu_e}^0/\Phi_{\bar\nu_x}^0$
decreases shown in the left panel of
Fig.~\ref{fig:parametric}. This is another feature depending on the adopted fluxes that was not previously found in the literature. Figure~\ref{fig:double-peak} refers to 
the uppermost and lowermost curves of the left panel of 
Fig.~\ref{fig:parametric}. We assume them as representative cases of a single-peaked (upper right panels)  or a double-peaked 
(lower right panels)
power spectrum, respectively.
The left panels of Fig.~\ref{fig:double-peak} show  the contributions of $\bar\nu_e$ and $\bar\nu_x$ to the observable signal $E^2 F_{\bar\nu_e}^{\oplus}$ while the right panels show  the different contributions to the power spectrum.
It is worthwhile to notice that the power-spectrum $G(k)$  of the signal [Eq.~(\ref{eq:contpower})]
 can be decomposed as  
\begin{equation}
G(k) = G_{{\bar\nu}_e}(k) + G_{{\bar\nu}_x}(k) + G_{{\bar\nu}_e {\bar\nu}_x}(k) \ ,
\label{eq:powerthree}
\end{equation}
where the first contribution on the right-hand-side is associated with $E^2 (1 - \bar{P}_{2e}) F^0_{\bar\nu_e}$, 
the second with $E^2 \bar{P}_{2e} F^0_{\bar\nu_x}$, and the third with the cross-correlation of the previous two. 
The three terms in the power-spectrum  are peaked at very similar frequencies.
Since  $G_{{\bar\nu}_e {\bar\nu}_x}(k)$  can assume
negative values, it can sometimes determine the appearance of the double-peak feature, especially
when the positive and negative  terms become comparable, as in the lower right panel of Fig.~\ref{fig:double-peak}.

\section{Neutrino detection}  
\label{sec:detectors}

In this Section we describe the main aspects and ingredients of 
our calculations  of supernova neutrino event rates.
The oscillated SN neutrino fluxes at the Earth, $F_\nu$,  
must be convolved with the differential cross section   
$\sigma_e$ for electron or positron production, 
as well as with the energy resolution function $R_e$ of the detector, 
and the efficiency $\varepsilon$ (that we assume equal to one above the energy threshold), in order to finally   
get observable event rates~\cite{FogliMega}: 
\begin{equation}  
N_e = F_\nu \otimes \sigma_e \otimes R_e \otimes \varepsilon\ .  
\label{Conv}  
\end{equation}  

We will now describe the main 
characteristics of four types of detectors we have used to calculate the signals in the presence of the Earth matter effects,
namely  water Cherenkov detectors, 
scintillation detectors, liquid Argon Time Projection Chambers, and ice Cherenkov detector Icecube.    

\subsection{Water Cherenkov detectors}

In large WC detectors, the dominant channel for supernova 
neutrino detection is the inverse beta decay of electron 
antineutrinos\footnote{We will neglect the 
subleading neutrino interaction channels in the detectors, 
assuming that they can be separated at least on a statistical basis.}
\begin{equation}
{\bar\nu}_e + p \to n+ e^+ \,\ .
\end{equation}
For this process, we take the differential cross section 
from~\cite{Strumia:2003zx}.  
The total cross section grows approximatively as $E^2$.
We fold the differential cross sections for $e^{+}$ production with
a Gaussian energy resolution function of width $\Delta$. 
The value of $\Delta$ is predominantly determined by the
photocathode coverage of the detector. 
For our calculations we assume  \cite{FogliMega}
\begin{equation}
\Delta_{\rm{WC}}/\textrm{MeV} = 0.47\sqrt{E_e/\textrm{MeV}} \,\ ,
\end{equation}
where $E_e$ is the true positron energy. 
We assume  as fiducial 
volume  400~kton~\cite{Autiero:2007zj}.

\subsection{Scintillation detectors}

In liquid SC detectors, the main channel for SN neutrino detection 
is the inverse beta decay of ${\bar\nu}_e$'s, the same as that in
WC. However, here the positrons are detected through 
photons produced in the scintillation material. Since a larger
number of photons can be produced in a SC detector, 
these have typically a much better energy resolution than the 
WC detectors. 
The energy resolution of the SC detectors is determined 
by the number of photo-electrons produced per MeV, which for 
this type of detectors is expected to be given by
as good as~\cite{Wurm:2007cy}
\begin{equation}
\Delta_{\rm{SC}}/\textrm{MeV} = 0.07\sqrt{{E_e}/\textrm{MeV}} \,\ .
\label{sc-resolution}
\end{equation}
Indeed, the energy resolution
of a SC detector may be better by more than a factor of  6 than a WC. 
Since the Earth matter oscillations described in the
previous section may get smeared out by the finite energy resolution of the
detector, it is clear that the energy resolution plays a crucial role 
in the efficiency of detecting Earth effects.
 For our studies, we assume a fiducial mass of  50~kton~\cite{Wurm:2011zn}.
  
\subsection{Liquid Argon Time Projection Chambers}  
  
LAr TPC detectors   would be particularly sensitive to SN electron 
neutrinos through their charged current interactions with Ar nuclei  
\begin{equation}  
 \nu_e + {}^{40}Ar \to {}^{40}K^{\ast} + e^- \,\ ,  
\end{equation}  
which proceed via the creation of an excited state of ${}^{40}K$ and its  
subsequent gamma decay. The Q-value for this 
inverse beta decay process is $1.505$~MeV. The cross-section 
for the charged current reaction is taken from \cite{Cocco:2004ac}.
The one     
for leptons in LAr TPC has been calculated by the 
ICARUS collaboration  which reports~\cite{GilBotella:2003sz}
\begin{equation}
\Delta_{\rm{LAr}}/\textrm{MeV} = 0.11\sqrt{{E_e}/\textrm{MeV}} + 
0.02\, E_e / {\rm MeV} \,\ .
\label{LAr-resolution}
\end{equation}
The fiducial volume for SN  neutrino detection is taken to be 
100~kton~\cite{GilBotella:2004bv}.   

\subsection{Icecube}

A galactic SN $\nu$ burst would be detectable in Icecube by a   sudden, correlated increase in the photomultiplier count rate on a timescale on the order of 10~s (see Ref.~\cite{Abbasi:2011ss} for a recent description).
  In its complete configuration and with its data acquisition system,  IceCube
has 5160 optical modules~\cite{Abbasi:2011ss} and about 3 Mton
effective detection volume, representing the largest current detectors
for SN neutrinos.
The SN neutrinos streaming through the antarctic ice interact mostly
through ${\overline \nu}_e +p\to n  + e^+$ reactions.
While
fine-grained detectors, like WC detectors, reconstruct individual neutrinos on
an event-by-event basis, IceCube only picks up the average Cherenkov glow of the ice. 
The detection rate is given by~\cite{Dighe:2003be,Serpico:2011ir}
\begin{equation}
{\cal R}_{\bar\nu_e}=\int_0^\infty d E\,F_{\bar\nu_e}\,E_{\rm rel}(E)\,\sigma(E)\,,
\end{equation}
 with $E_{\rm rel}(E)$ being the energy released by a neutrino of energy $E$ and $\sigma(E)$ the 
inverse beta-decay cross section.
All other detector parameters (angular acceptance range, average quantum efficiency, number of useful Cherenkov photons per deposited neutrino energy unit,  average lifetime of Cherenkov photons, effective photo cathode detection area) have been fixed to the fiducial values adopted in~\cite{Dighe:2003be}, 
 to which we address to for further details.

\section{Detecting Earth matter effect}

\subsection{Single detector}

We present our results about the detectability of the Earth effect.  For our numerical calculations, we assume 
the mass-mixing parameters as in Eqs.~(\ref{masses}--\ref{theta1312}).  Note that, although we stick to the best fit values of the 
most recent $3 \nu$ global analysis, our conclusions do not qualitatively change for small variations of the adopted numerical values of the mixing parameters.
 We will also assume that the path-length crossed by neutrinos in the Earth is 
 $L=6000$~km. 
 
\begin{figure}[ht]

  \begin{minipage}[b]{0.32\linewidth}
	\centering 10 kpc\\
	\includegraphics[width=\textwidth]{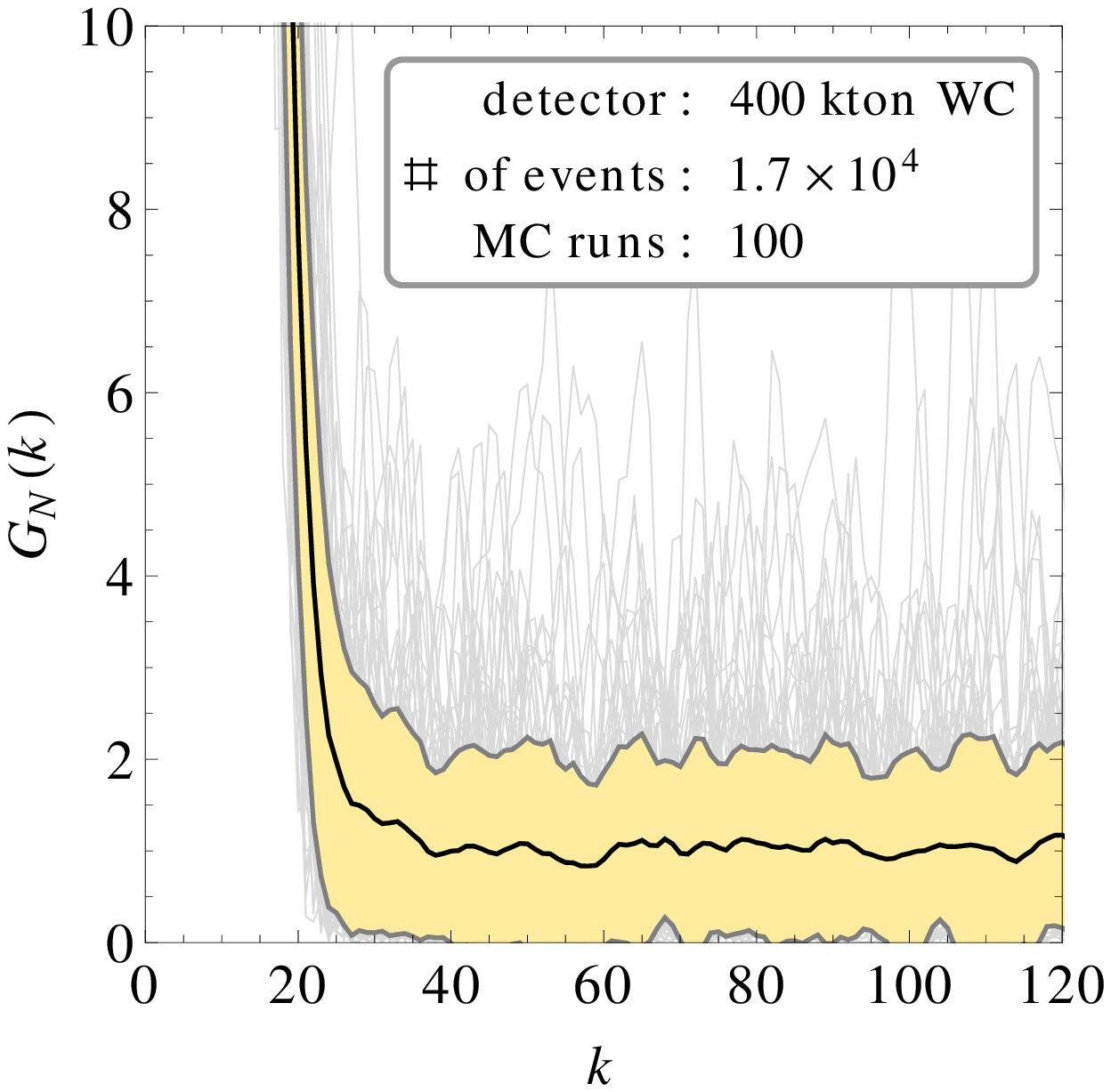}
  \end{minipage}
  \begin{minipage}[b]{0.32\linewidth}
	\centering 1 kpc \\
	\includegraphics[width=\textwidth]{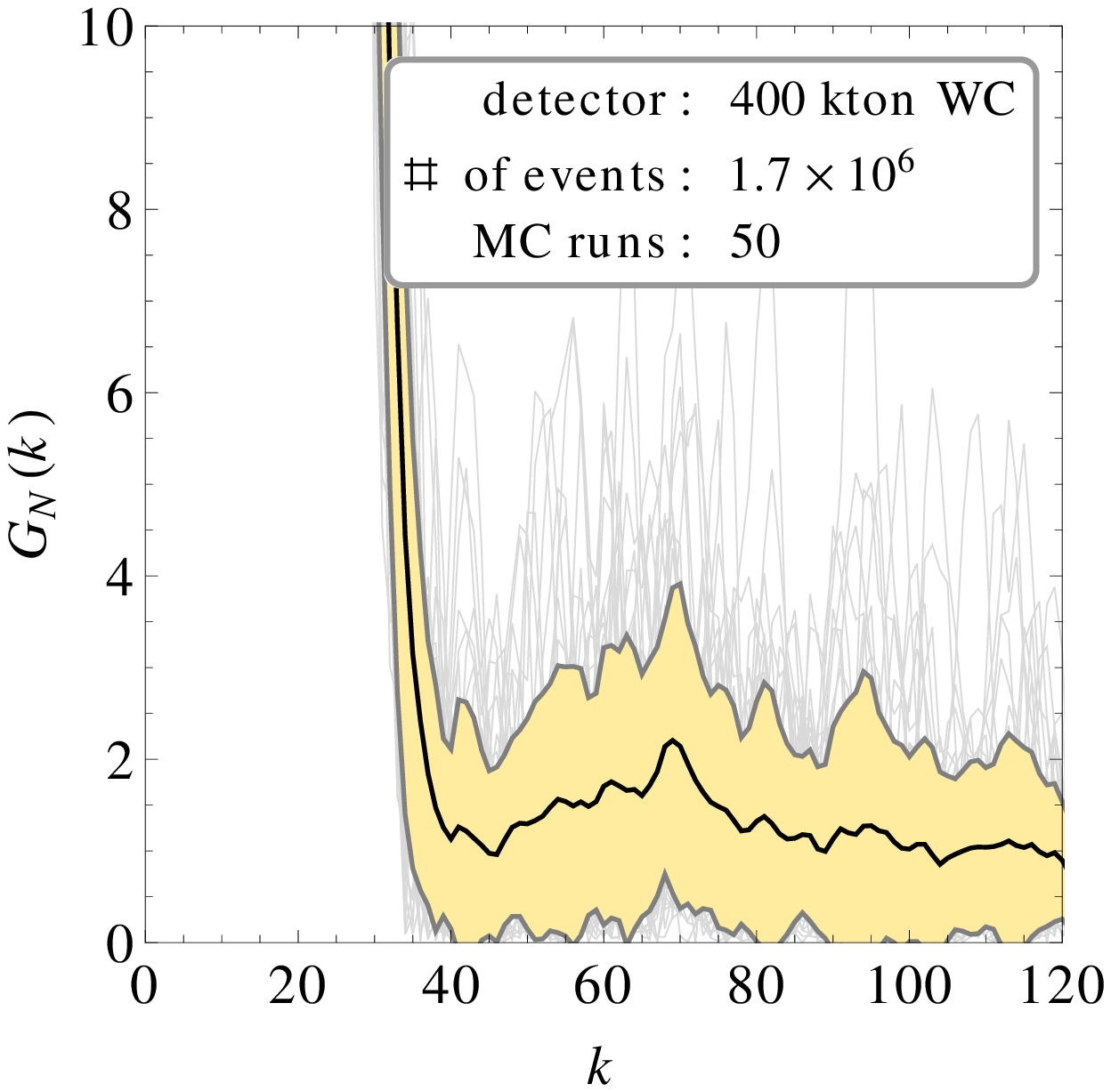}
  \end{minipage}
  \begin{minipage}[b]{0.32\linewidth}
	\centering 0.2 kpc \\
	\includegraphics[width=\textwidth]{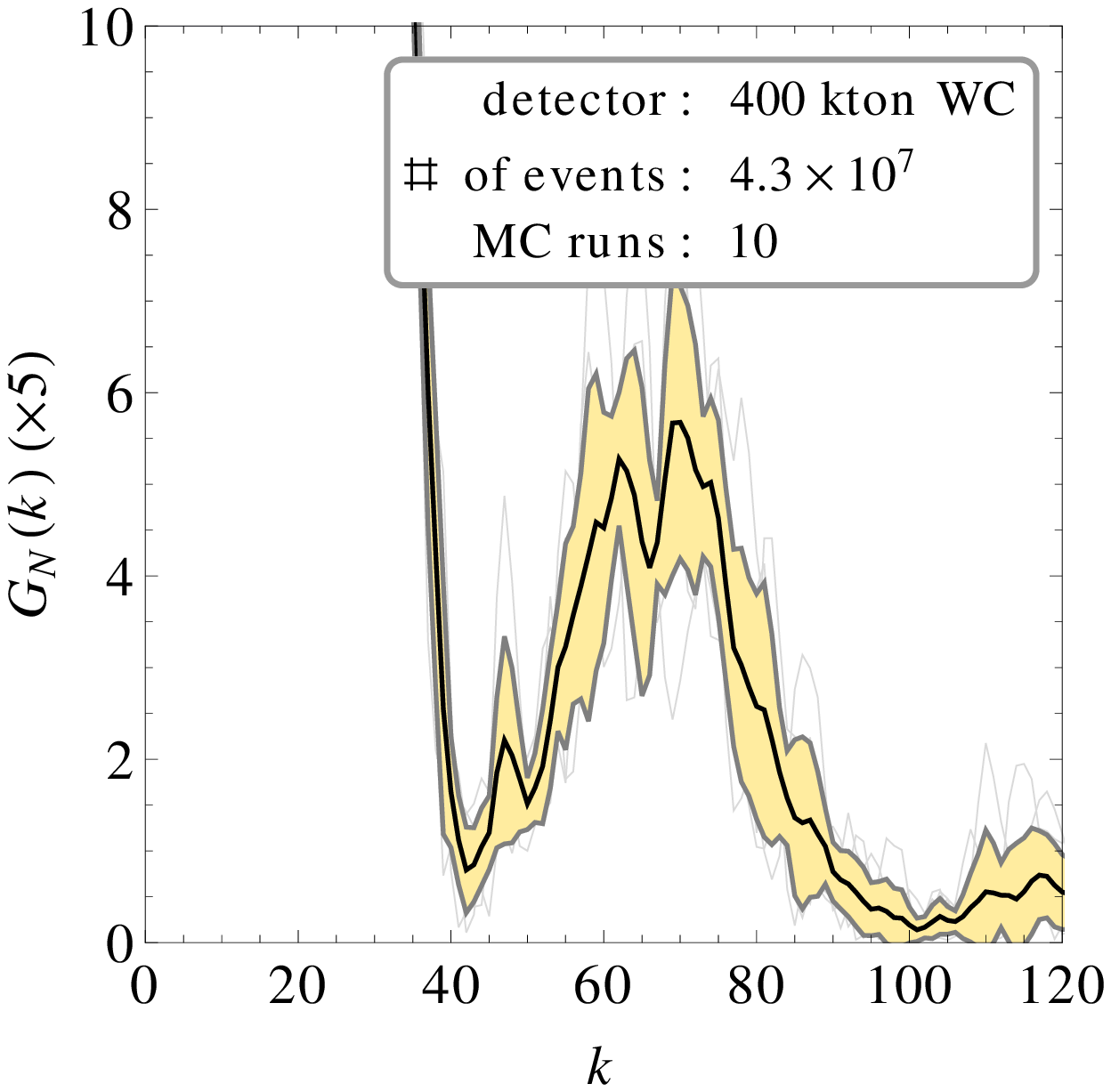}\\[4mm]
  \end{minipage}
  \begin{minipage}[b]{0.32\linewidth}
	\centering 
	\includegraphics[width=\textwidth]{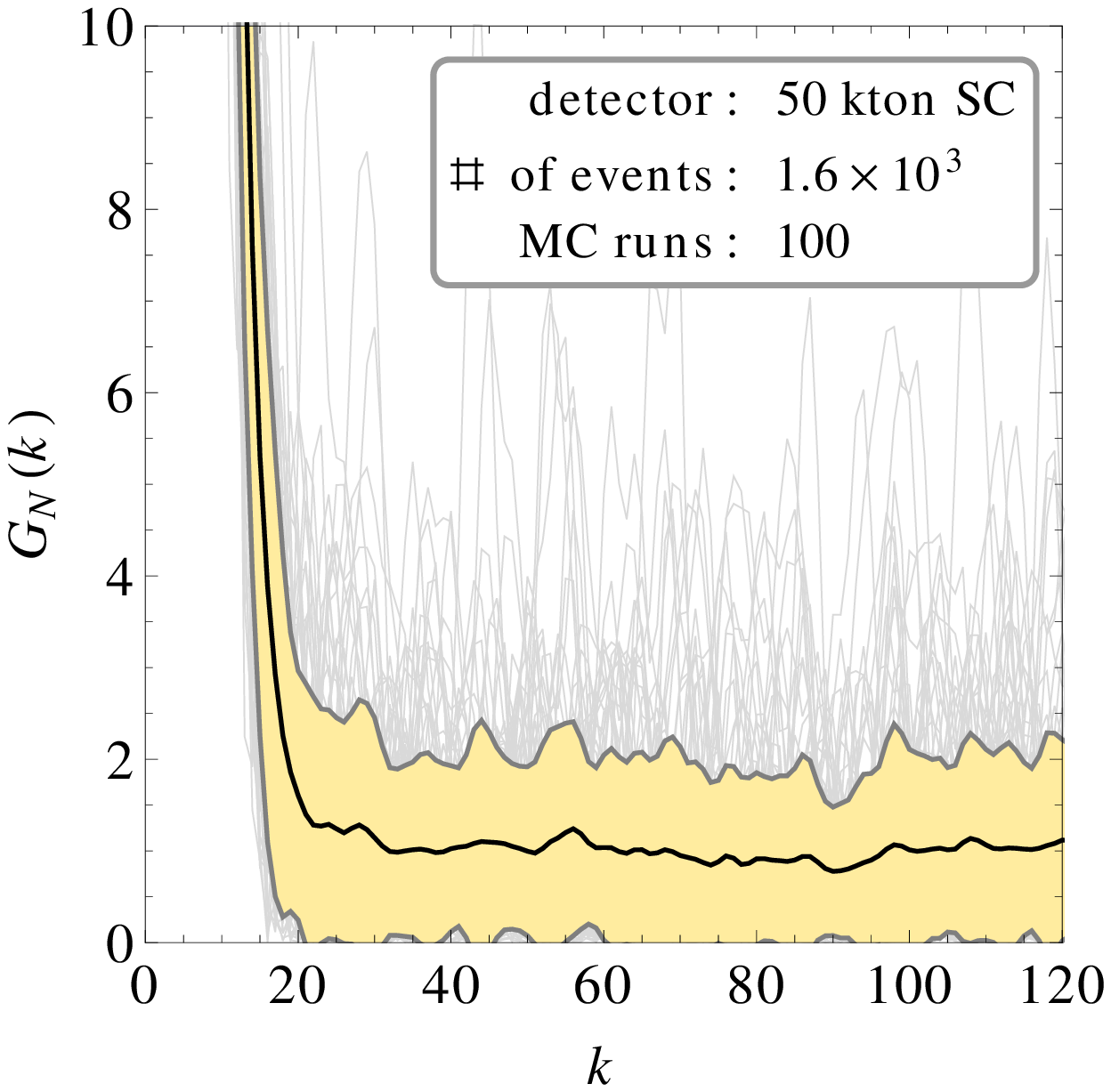} 
  \end{minipage}
  \begin{minipage}[b]{0.32\linewidth}
	\centering 
	\includegraphics[width=\textwidth]{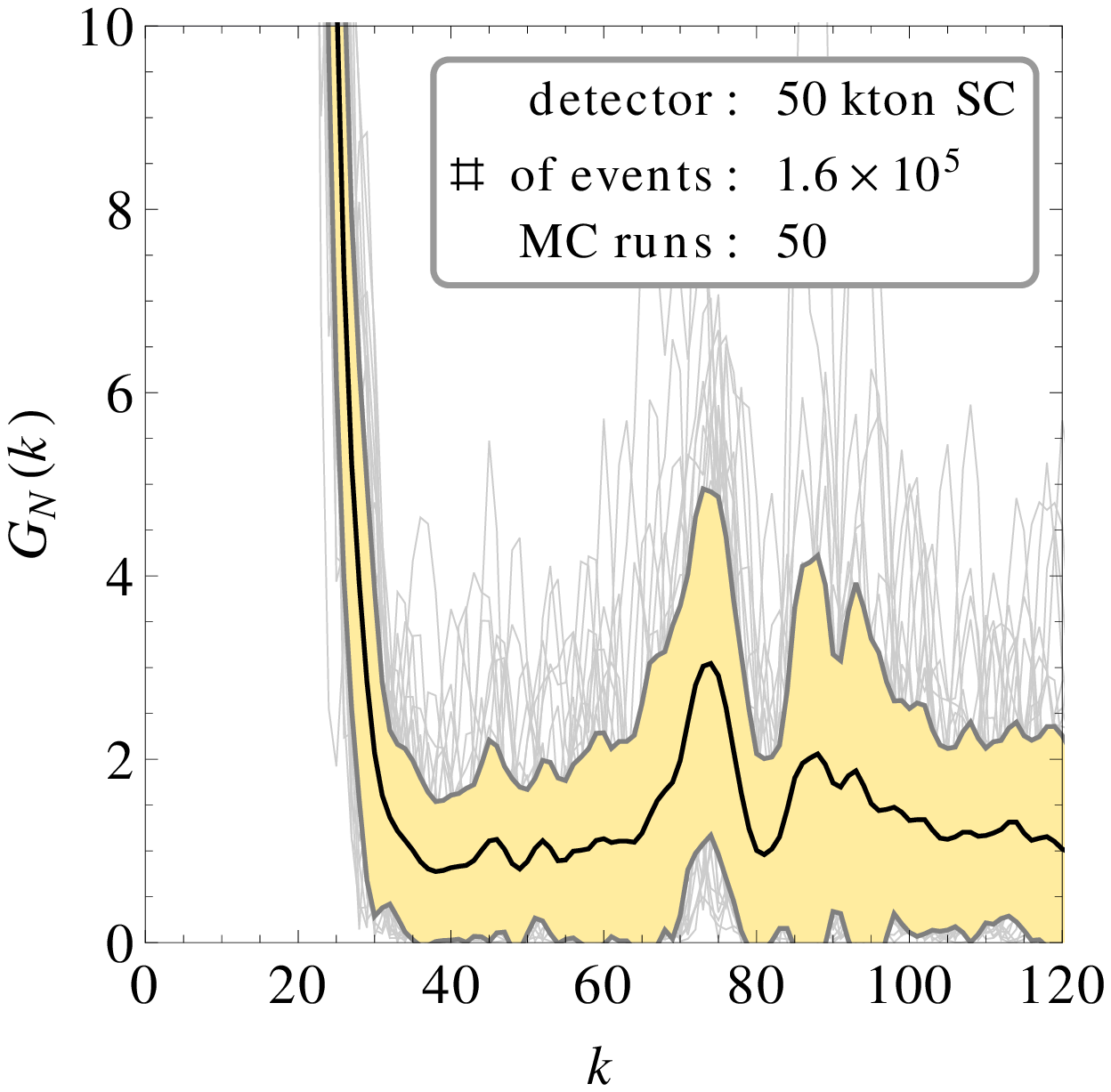}  
  \end{minipage}
  \begin{minipage}[b]{0.32\linewidth}
	\centering 
	\includegraphics[width=\textwidth]{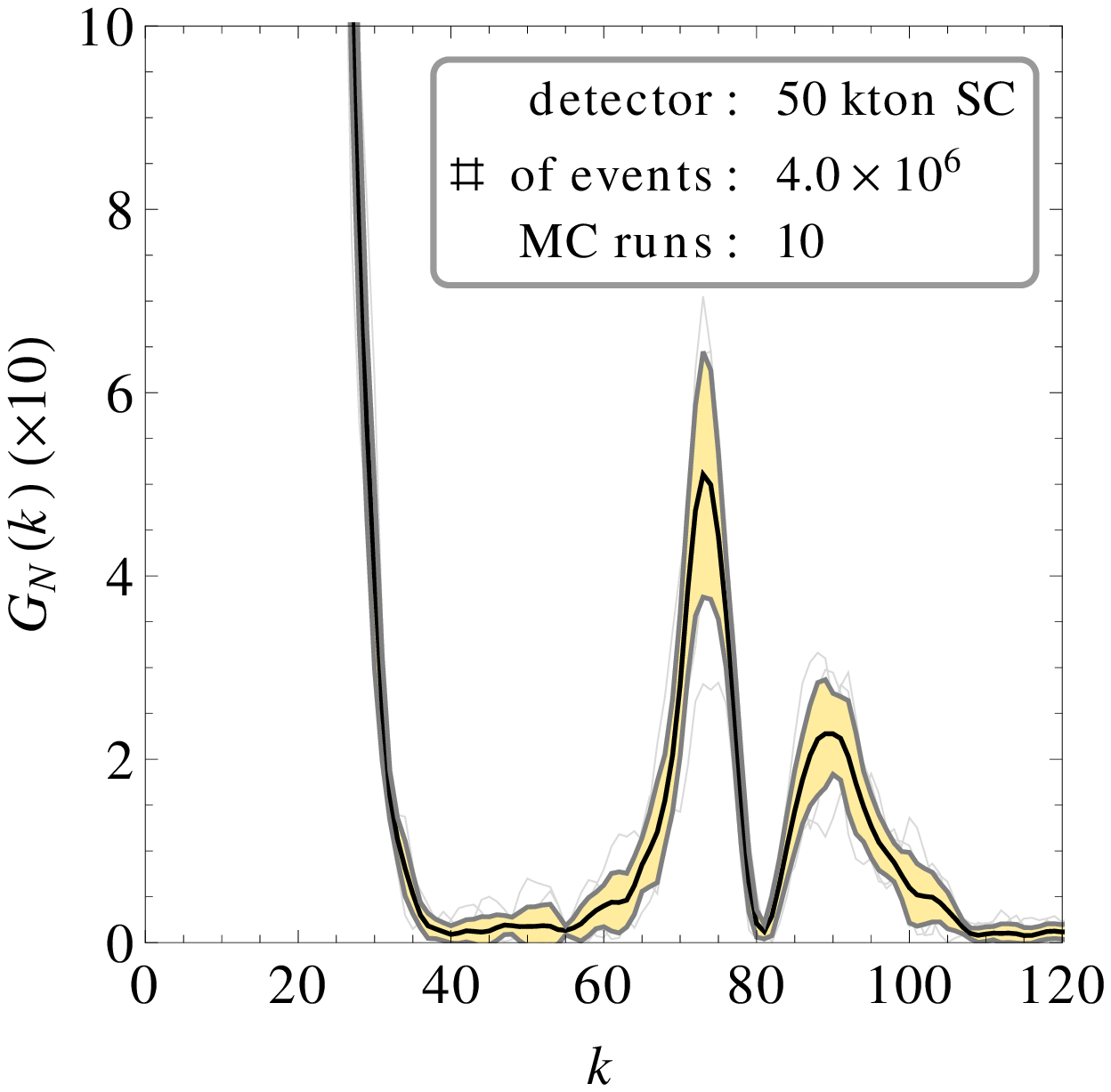} 
  \end{minipage}
  \begin{minipage}[b]{0.32\linewidth}
	\includegraphics[width=\textwidth]{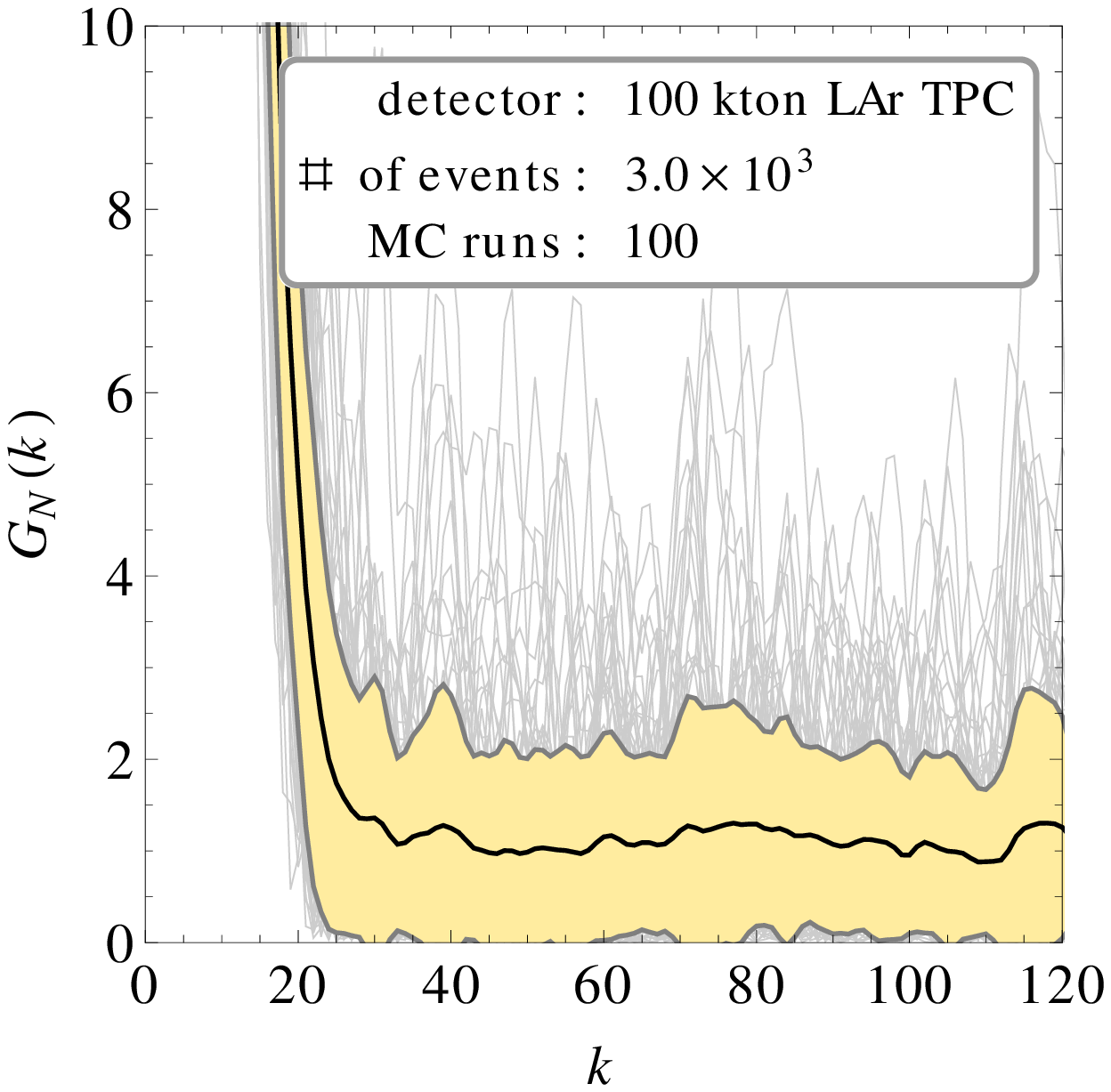}
  \end{minipage}
  \begin{minipage}[b]{0.32\linewidth}
	\includegraphics[width=\textwidth]{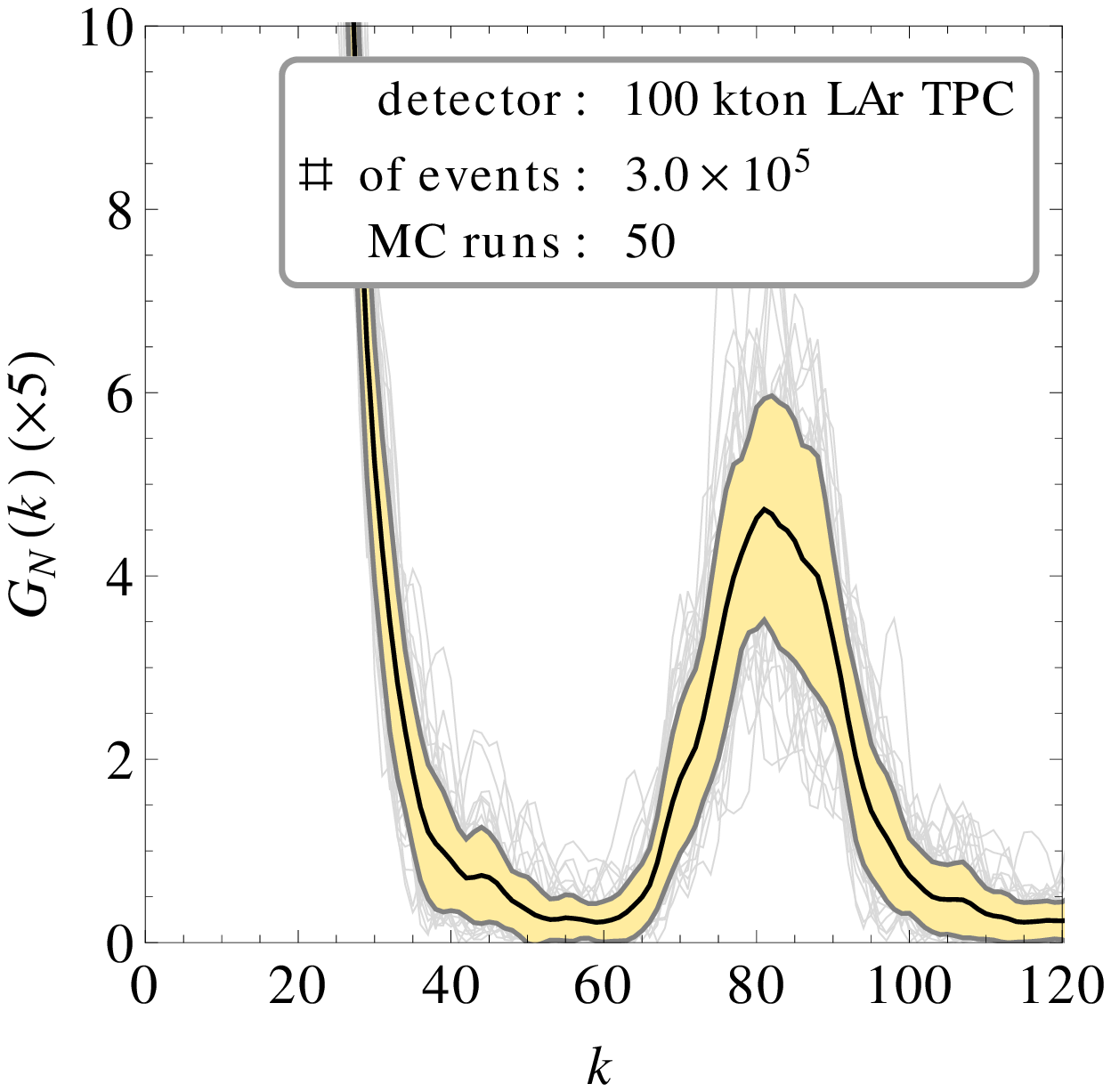}
  \end{minipage}
  \begin{minipage}[b]{0.32\linewidth}
	\includegraphics[width=\textwidth]{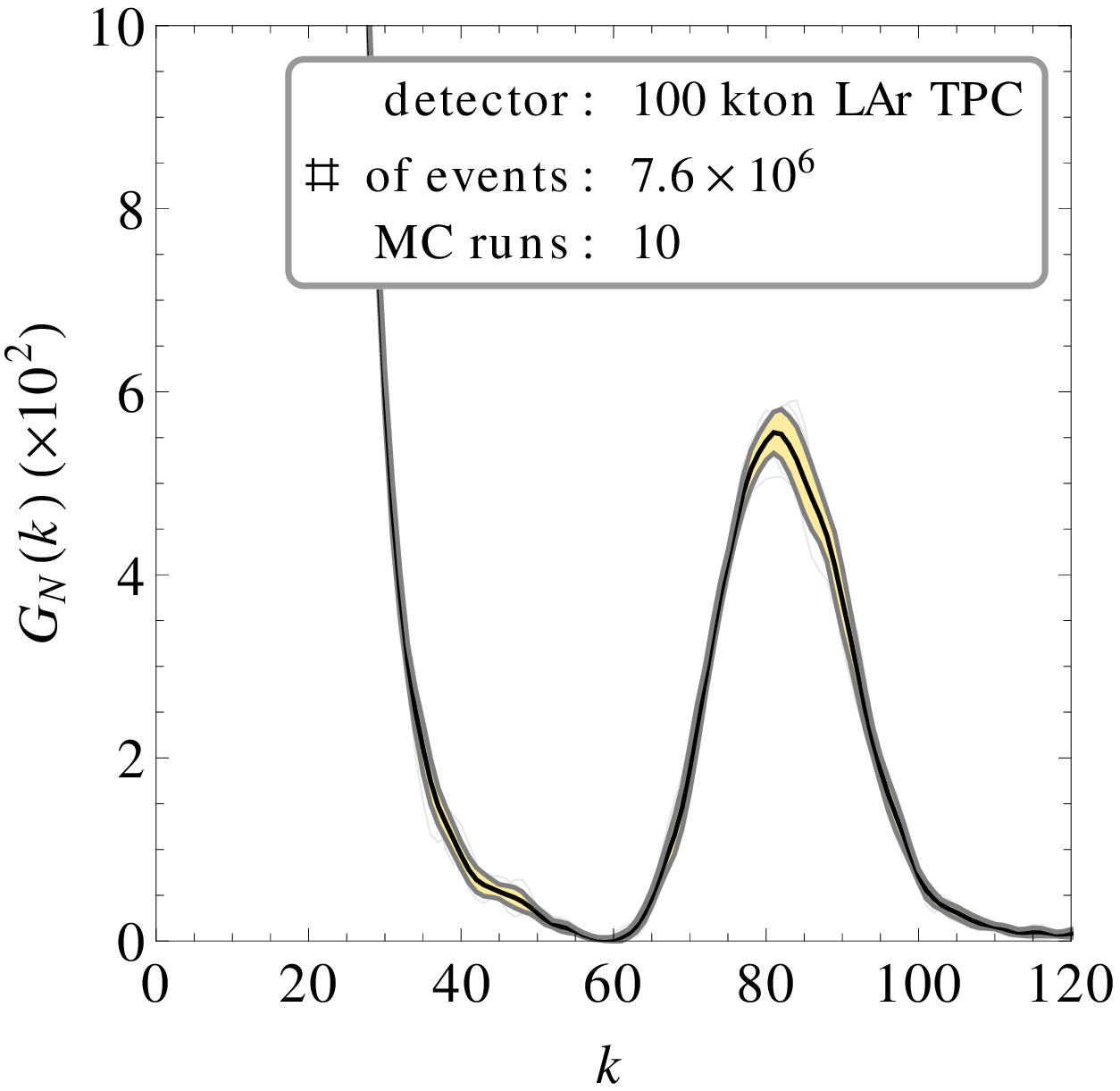}
  \end{minipage}
  \caption{\label{fig3}Power spectrum $G_N(k)$ of the Earth matter effect for a Galactic SN at $d=10$~kpc (left panels),
 $d=1$~kpc (central panels) and $d=0.2$~kpc (right panels). The upper panels refer to a 400~kton WC
detector, the middle panels to a 50~kton SC, and the lower panels to a 100~kton LAr TPC.
The light curves corresponds to different MonteCarlo realizations of the  power spectrum $G_N(k)$, the thick curve
to the average over the different realizations, and the band to the $\pm 1 \sigma$ variance level.}
\end{figure}

 Figure~\ref{fig3} shows the power-spectrum $G_N(k)$
[defined as in Eq.~(\ref{power})]
of the SN $\nu$ signal in the WC (upper panels), SC (central panels) and LAr TPC detectors
(lower panels).
We discuss the results for three different SN distances from Earth ($d=10,1,0.2$~kpc). 
 Since the Earth matter probability is time-independent,
we consider as neutrino signal the time-integrated rate during the accretion phase taken  from   the 15 $M_{\odot}$  Garching SN simulation  (see Sec.~2 and  Table~\ref{tab:fluxes} for the time-integrated flux parameters). We produce
 10, 50 or 100 realizations of
the SN neutrino spectra at the Earth via Montecarlo simulations.
The thin gray lines  in Fig.~\ref{fig3} correspond to different realizations of 
 $G_N(k)$ [Eq.~(\ref{power})].  The thick black line corresponds to the  power-spectrum averaged over the different realizations $ G_N(k)$.
The light band around the average corresponds to the $\pm 1\sigma$ level. 

As it was already pointed out in \cite{Dighe:2003jg}, the frequency
range at which the peak (or peaks) of $G_N(k)$ has to be expected can be
predicted in advance. But peaks at different values of $k$ are also
visible (thin gray lines); their positions depend on the discrete energy
spectrum of the $N$ detected neutrinos and thus
both on input fluxes and on stochastic, finite statistics effects. When
the Earth modulation effect is not sizable enough, its associated peak
can be comparable to or shadowed by these features. In order to quantify
the observability of the Earth effect peak, we compare the mean value of
$G_N(k)$ to the expected noise ($\sim 1$). The fact that they are
compatible at $1 \sigma$ means that the Earth effect will not produce a
visible peak in at least the 68~\% of cases. This situation is clearly
visible in some of the plots of Fig.~\ref{fig3} (e.g. WC and SC at 1
kpc) in which the mean value of $G_N(k)$ is always compatible with a
null detection at $1 \sigma$, despite the fact that some of the
realizations exhibit a very well defined Earth effect peak.

Starting with a typical SN at $d=10$~kpc (left panels of Fig.~\ref{fig3}),  $G_N(k) \simeq 1$ for $k\gtrsim 40$, for all the three detectors
as expected
in the absence of Earth modulation.
No peak in $G_N(k)$ associated with the Earth effect is visible. 
For a SN at $d=1$~kpc, a peak around $k\simeq 70$ seems to emerge in the average $G_N(k)$ for the WC and for the LS. 
However,  the power spectrum
is compatible with 1 at $ 1\sigma$, i.e. with the expectations without the Earth crossing.
Conversely, in the case of LAr TPC the peak in $G_N(k)$ is clearly visible at $k\simeq 80$. 
Indeed, this detector has the benefit of testing the Earth effect in the neutrino channel where
the difference in the average energies/fluxes of $\nu_e$ and $\nu_x$ is larger than in the antineutrino sector. 
Therefore, the Earth signature is enhanced.
We checked that  the Earth effect starts to  be visible in a
LAr TPC at a few kpc for our benchmark SN neutrino fluxes. 
In the right panels of Fig.~\ref{fig3},  we show the power spectrum for the lucky case of a  very close-by
SN at $d=0.2$~kpc.  The peak in the power spectrum is clearly visible in the three different
detectors.
In particular, in the case of a liquid SC the superior energy resolution allows to see the double-peak 
structure in the power-spectrum discussed in Sec.~3. 

We repeated the same analysis with other eight SN models from the Garching simulations with 
progenitor mass between 12 and 40 $M_{\odot}$ (see Fig.~1 in~\cite{Serpico:2011ir}) finding similar
results to what shown before. Also for the accretion phase of the Basel/Darmstadt progenitors with
 10.8~$M_{\odot}$ and 18~$M_{\odot}$~\cite{Fischer:2009af}, our results are similar. 
Finally, we calculated the power-spectrum in  presence of 
the Earth matter also during the cooling phase for the Basel/Darmstadt simulations, characterizing $\nu$ oscillations as described in 
Sec.~3. We find that, since during this phase the spectral differences among different flavors are 
very small, no peak in the power spectrum appears even in the optimistic case of a very close-by SN (results not shown here). 

 As anticipated in the previous sections,  the recent
supernova simulations point towards mean energies that are lower than
previously considered in the literature and and fluxes that are closer among themselves during the cooling phase. As evident from Fig.~\ref{fig3}, these spectral features 
generate a destructive
interference in the power spectrum sometimes responsible for the appearance of a double peak  and  suppress the intensity of the expected Earth peak (see also Figs.~\ref{fig:parametric},\ref{fig:double-peak}). Moreover, we take into account the time-dependence of the neutrino fluxes and discuss a time-integrated analysis for both the accretion and the cooling phase. Therefore an eventual enhancement of the power spectrum peak due to a lucky  choice of the 
spectral parameters (i.e., a choice of the initial fluxes corresponding to 
a particular post-bounce time, as in the existing literature) could be averaged out with the time integration. Of course, despite the major improvements on the simulations side over the last decade (e.g. on dimensionality, weak interaction physics, general relativity, etc.) one cannot exclude that the current supernova paradigm is oversimplified and forthcoming  supernova simulations, including effects not yet considered  (or new physics), might point towards an enhancement of the differences among the neutrino fluxes, allowing a better resolution for the Earth peak.

\subsection{Two detectors}\label{2det}

The Earth effect could produce a modification in the  SN $\bar\nu_e$ light-curve
measured by Icecube. Therefore, together with a high-statistics Mton WC detector
it could detect the Earth effect (if only one of the two detectors is shadowed)
by the relative  difference in the temporal signals~\cite{Dighe:2003be}.

\begin{figure*}[!]
\epsfig{file=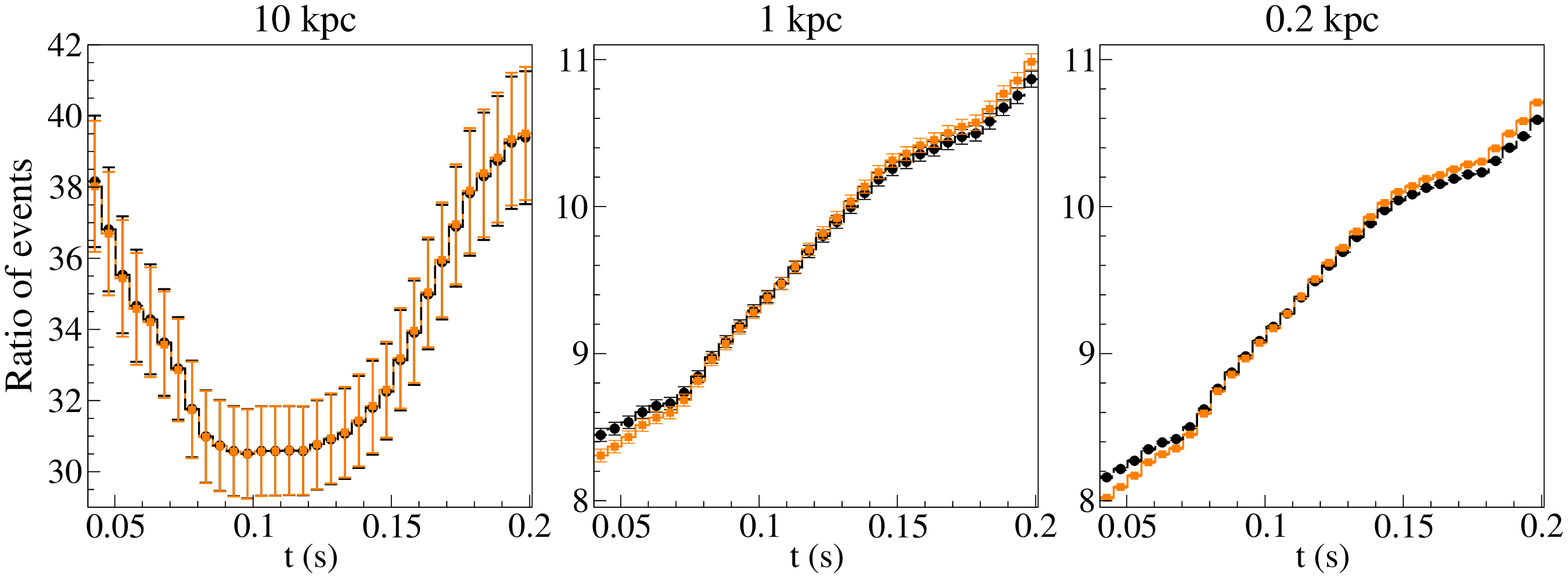,width=0.4\hsize,angle=270}
\caption{\label{fig-IceCube}
Events ratio in  IceCube with respect to a 400  kton WC detector. The
Mton detector is always considered un-shadowed by the Earth. 
The case when  IceCube is shadowed is shown
with the  orange squares, whereas the case of Icecube un-shadowed 
plotted by  black circles. The left panel  represents
the events ratio for a SN at 10 kpc, the central at 1 kpc, and the right  at 0.2 kpc, respectively.}
\end{figure*}

Figure~\ref{fig-IceCube} shows the ratio of the counting rate for IceCube and  for a 400 kton WC detector, 
taking  the input for the un-oscillated $\nu$ fluxes  from the accretion phase of the  15 $M_{\odot}$  Garching model. 
 The counting rate in IceCube also includes the noise  from the photomultipliers
(280 Hz per optical module). 
We assume that the WC detector is un-shadowed. In the case that 
 IceCube is  shadowed we indicate the ratio of events with  squares,
while when Icecube is un-shadowed the ratio   is plotted with  circles. The left panel  represents the ratio for a SN at 10 kpc, the central  at 1 kpc, and the right at 0.2 kpc, respectively. In all the cases, we employ the time-dependent values of the parameters $L_\nu$, $\langle E_\nu\rangle$ and $\alpha_\nu$ as obtained from 15 $M_{\odot}$  Garching model. Note also that the shape of the ratio changes with the distance of the SN, since the measured signal in Icecube is the sum of a 
time-independent background rate (independent of the distance) and a true SN lightcurve whose normalization depends on the distance.

The left panel clearly shows that the ratio with and without Earth matter effects are nearly inseparable for a SN at $d=10$~kpc, since   
the statistical errors  are  larger than the difference.
For a SN at $d=1$~kpc, the ratio in the shadowed and un-shadowed case still have too large statistical errors 
to be clearly separated. Conversely, the shadowed and un-shadowed cases are statistically separable at $d=0.2$~kpc. 
However, since the differences between the two curves are relatively small  (maximum difference $\sim 1.7 \%$ at $40$~ms) and the  the initial neutrino fluxes  are not known with such precision, it is unlikely that the ratio of events could be used to diagnose the Earth effect,  unless the comparison can be performed
 between two {\it very similar} detectors, with systematics canceling out. Otherwise, moderate variations in the  distance of the SN events or the time-dependence used for the parameters ($L_\nu$, $\langle E_\nu\rangle$ and $\alpha_\nu$) may alone alter the ``reference''  curve with respect to which compared an eventual signal.

\section{Conclusions}

The Earth matter effect in supernova neutrinos would be  an interesting tool to diagnose
the neutrino mass hierarchy at ``large''   $\theta_{13}$. 
Motivated by the recent measurement of this angle~\cite{An:2012eh,Ahn:2012nd} and by
the vivid discussion  for future large underground neutrino detectors,
we found worthwhile to reevaluate the chance to detect this effect in future neutrino experiments.
In order to achieve a realistic characterization of this signature, we adopted state-of-the-art SN simulation inputs~\cite{Fischer:2009af,Serpico:2011ir}
to describe the un-oscillated SN neutrino signal. 
The detection of the modulation in the neutrino spectra induced by the Earth crossing
largely  depends on the neutrino  average energies  and on
 the flavor-dependent differences between the primary spectra. At this regard,  recent supernova simulations  indicate lower average energies
than previously expected~\cite{Fischer:2009af,Huedepohl:2009wh,Serpico:2011ir} and a tendency towards the  equalization of the neutrino fluxes of different flavors during the cooling phase~\cite{Fischer:2011cy}. This makes the detection of the Earth matter effect
 more challenging than what assumed in previous works based on  outdated SN simulations. 
 
In order to diagnose the modulation in the SN $\nu$ energy spectrum induced by the Earth
crossing, we perform a Fourier analysis of the neutrino signal in a 400 kton WC detector, 
in a 50 kton liquid SC and in a 100 kton LAr TPC. 
For all these  detectors, we  found that coming from a typical galactic SN at $d=10$~kpc,
no signature of the Earth matter effect is observable in the measured SN neutrino burst. 
Moreover, also in the more  optimistic case of a  close-by SN at $d=1$~kpc, the chances
to detect the  Earth matter
signature appear statistically weak in the antineutrino signal. Conversely, a signal would show up
in the $\nu_e$ signal detectable at a LAr TPC. Only for relatively close stars which might evolve into core-collapse supernovae at unpredictable
future times like Betelgeuse, Mira Ceti, and Antares (at $d\lesssim 0.2$~kpc), the Earth matter
signal would be detectable with high significance in both neutrino and antineutrino signals. 
Finally, Icecube taken as co-detector to monitor the Earth effect together with a Mton WC detector is not able to detect any sizeable variation in the SN neutrino event rate associated with 
the Earth matter effect for any galactic supernova. 

These new results based on the state-of-art SN simulations dramatically change the 
previous perspectives of detection of the Earth matter effect with supernova neutrinos based on
an  outdated choice for the primary SN neutrino fluxes, as reported from previous supernova simulations. 
In particular, Mton WC detectors and large liquid scintillators would be able to observe the 
Earth matter signature only for very-close by (and rare) SNe,  provided that the electronics of the detector will
be able to cope with huge rates of events.
A 100 kton LAr TPC which  starts to monitor the Earth signature from SNe at a distance of
few kpc from the Earth, would statistically have $\sim 10$\% of chance  to see this signature
from the next galactic supernova explosion~\cite{Mirizzi:2006xx}.

As a consequence of our finding, the possibility to  determine the neutrino mass hierarchy with the next galactic SN neutrino burst 
requires to be rediscussed.  Of course, a caveat is that, while current SN simulations have improved in many ways
with respect to  one or two decades ago, the results obtained should still be considered as indicative, and
an empirical test would certainly be welcome. Turning the argument around, we can say that---barring an exceptional situation of a close-by SN---a positive detection of
the Earth matter effect in future data would come as a surprise and probably invalidate the current models. A negative result most likely would
turn into constrains in the flavor flux difference vs. average energy differences parameter space,  thus indirectly testing these models.

Of course, this should not discourage experimentalists to devote efforts to achieve a detailed
measurement of the neutrino flux and spectra from a Galactic supernova; it would be ``per se''
a bonanza for testing the astrophysical models and the detailed understanding of the core-collapse SN mechanism.

If Earth matter effect in SNe is now not so promising as thought before, there 
are still other intriguing signatures in the SN neutrino signal that could give important information on the mass 
hierarchy. 
In particular, the early $\nu_e$ neutronization peak that could be detected in a Mton WC detector or 
in a large LAr TPC would provide a clear signature to extract the neutrino mass hierarchy~\cite{Kachelriess:2004ds}.
Also the early signal  rise during the accretion phase, detectable by all the detectors discussed in this work,
could encode an imprint of the neutrino mass ordering~\cite{Serpico:2011ir}. 
In conclusion, supernova neutrinos still represent a unique astrophysical probe of neutrino physics and
astrophysics under extreme conditions.

\section*{Acknowledgements} 
We are grateful to Tobias Fisher, Thomas Janka and collaborators for providing the supernova data.
E.B. and I.T. thank  Georg Raffelt for 
useful discussions.
The work of E.B.  was supported by LAGUNA-LBNO. 
The work of S.C. and A.M. was supported by the German Science Foundation (DFG)
within the Collaborative Research Center 676 ``Particles, Strings and the
Early Universe''. 
 I.T. acknowledges support from the Alexander von Humboldt Foundation,
  partial support from the Deutsche Forschungsgemeinschaft under grant EXC-153 and by the  European Union FP7 ITN
INVISIBLES (Marie Curie Actions, PITN-GA-2011-289442).

\section*{References} 


\end{document}